\documentclass[
  final,
  1p,
  10pt
]{elsarticle}

\usepackage{iftex}
\usepackage[hidelinks]{hyperref}
\usepackage{mathtools}
\ifLuaTeX
  \IfFileExists{phiso.sty}{
    \usepackage{phiso}
    \def\vecuAst{\vec{u}^\ast}
    \def\opLrho{\operatorname{\mathup{L}}_\rho}
    \def\opKrho{\operatorname{\mathup{K}}_\rho}
    \def\opPrho{\operatorname{\mathup{P}}_\rho}
    \def\opV{\operatorname{\symbfup{V}}}
    \def\opVrho{\opV_\rho}}{}
\else
  \usepackage{amsmath}
  \usepackage{isomath}
  \usepackage{upgreek}
  \usepackage{derivative}
  \renewcommand\vec{\mathbfit}
  \newcommand\mat{\mathbfit}
  \newcommand{\iu}{\mathrm{i}\mkern1mu}\let\imath\iu
  \def\opV{\operatorname{\mathbf{V}}}
  \def\opVrho{\opV\kern-0.4em_\rho}
  \def\opLrho{\operatorname{\mathrm{L}}_\rho}
  \def\opKrho{\operatorname{\mathrm{K}}_\rho}
  \def\opPrho{\operatorname{\mathrm{P}}\kern-0.3em_\rho}

  \newcommand{\trans}{^{\mkern-1.5mu\mathsf{T}}}
  \newcommand{\Order}{\operatorname{O}}
  \newcommand{\Frou}{\operatorname{\mathit{F\kern-.07em r}}}
  \newcommand{\Reyn}{\operatorname{\mathit{R\kern-.04em e}}}
  \newcommand{\Schm}{\operatorname{\mathit{S\kern-.07em c}}}

  \AtBeginDocument{
    \let\nablaSymbol\nabla
    \renewcommand{\nabla}{\boldsymbol{\nablaSymbol}}
  }
\fi
\def\vecuAst{\vec{u}^\ast}
\newcommand{\ugrad}[1]{\vec{u}\cdot\nabla#1}
\makeatletter
\@ifundefined{odif}{\newcommand\odif\mathrm{d}}{}
\makeatother

\usepackage{caption}
\usepackage{subcaption}
\usepackage[noabbrev]{cleveref}
\graphicspath{{figures/}{build/}}
\usepackage{pgfplots}
\pgfplotsset{
  compat=newest,
  colormap/inferno/.style={%
    /pgfplots/colormap={inferno}{%
      rgb=(0.001462, 0.000466, 0.013866)
      rgb=(0.037668, 0.025921, 0.132232)
      rgb=(0.116656, 0.047574, 0.272321)
      rgb=(0.217949, 0.036615, 0.383522)
      rgb=(0.316282, 0.053490, 0.425116)
      rgb=(0.410113, 0.087896, 0.433098)
      rgb=(0.503493, 0.121575, 0.423356)
      rgb=(0.596940, 0.154848, 0.398125)
      rgb=(0.688653, 0.192239, 0.357603)
      rgb=(0.775059, 0.239667, 0.303526)
      rgb=(0.851384, 0.302260, 0.239636)
      rgb=(0.912966, 0.381636, 0.169755)
      rgb=(0.956852, 0.475356, 0.094695)
      rgb=(0.981895, 0.579392, 0.026250)
      rgb=(0.987464, 0.690366, 0.079990)
      rgb=(0.973088, 0.805409, 0.216877)
      rgb=(0.947594, 0.917399, 0.410665)
      rgb=(0.988362, 0.998364, 0.644924)
    },
  },
  colormap/inferno,
  kappa axis/.style={%
    ylabel=$\kappa$,
    legend cell align={left},
    legend style={at={(0.90, 0.5)}, anchor=east},
    cycle list={
      only marks, every mark/.append style={fill=gray}, mark=*\\
      solid,mark=none\\
      only marks, every mark/.append style={fill=gray}, mark=square*\\
      dashed,mark=none\\
      mark=triangle*\\
    }
  },
  conv axis/.style={%
    ylabel={Residual},
    xlabel={Iteration number},
    xtick={0,20,...,100},
    xmax={110},
    legend style={at={(0.5, -0.2)}, anchor=north, legend columns=-1},
    unbounded coords=jump,
    cycle list={
      every mark/.append style={fill=gray}, mark=*\\
      every mark/.append style={fill=gray}, mark=oplus*\\
      every mark/.append style={fill=gray}, mark=otimes*\\
      every mark/.append style={fill=gray}, mark=pentagon*\\
      every mark/.append style={fill=gray}, mark=diamond*\\
      every mark/.append style={fill=gray}, mark=triangle*\\
    }
  },
  contour axis/.style={%
    view={0}{90},
    tick style={draw=none},
    ytick=\empty,
    xtick=\empty,
  }
}
\usepgfplotslibrary{groupplots}
\makeatletter
\def\IfTikzLibraryLoaded#1{%
  \ifcsname tikz@library@#1@loaded\endcsname
    \expandafter\@firstoftwo
  \else
    \expandafter\@secondoftwo
  \fi
}
\makeatother
\IfTikzLibraryLoaded{external}{\tikzexternalize[mode=list and make]}{}

\begin{document}


\begin{frontmatter}
\journal{Computers and Fluids}
\title{%
  A preconditioning for the spectral solution of
  incompressible variable-density flows%
}

\author[label1]{L. Reynier\corref{cor1}}
\ead{loic@loicreynier.fr}
\author[label1]{B. Di Pierro}
\author[label1]{F. Alizard}
\author[label2]{A. Cadiou}
\author[label2]{L. Le Penven}
\author[label1]{M. Buffat}
\cortext[cor1]{Corresponding author}
\address[label1]{%
  Université Claude Bernard Lyon 1,
  LMFA, UMR5509,
  69622 Villeurbanne, France
}
\address[label2]{%
  École Centrale de Lyon,
  CNRS, LMFA, UMR5509,
  69130 Écully, France
}

\ifLuaTeX
  \hypersetup{
    pdfauthor={Loïc Reynier},
    pdfkeywords={%
      Variable-density flows, %
      incompressible Navier-Stokes equations, %
      direct numerical simulation, %
      preconditioning, %
      elliptic solver%
    }
  }
\else
  \hypersetup{
    pdfinfo={
    pdfproducer={},
    Title={},
    Subject={},
    Author={},
    }
  }
\fi

\begin{abstract}
In the present study,
the efficiency of preconditioners for solving linear systems associated with
the discretized variable-density incompressible Navier--Stokes equations
with semi-implicit second-order accuracy in time
and spectral accuracy in space
is investigated.
The method,
in which the inverse operator for the constant-density flow system
acts as preconditioner,
is implemented for three iterative solvers:
the General Minimal Residual,
the Conjugate Gradient and
the Richardson Minimal Residual.
We discuss the method,
first, in the context of the one-dimensional flow case
where a top-hat like profile for the density is used.
Numerical evidence shows that the convergence is significantly improved
due to the notable decrease in the condition number of the operators.
Most importantly,
we then validate the robustness and
convergence properties of the method on two more realistic problems:
the two-dimensional Rayleigh--Taylor instability problem
and the three-dimensional variable-density swirling jet.
\end{abstract}


\end{frontmatter}

\section{\label{sec:intro}Introduction}

Flows with large spatial density variations play an important role
in widespread industrial and environmental applications such as fluid mixer
found in pharmaceuticals processes or pollutant dispersion phenomenon.

Especially,
the generation of baroclinic vorticity
due to the interaction of non-parallel pressure and density gradients,
and mass diffusion effects yield to a large variety of scale motions
that presents a significant numerical challenge.
In this context,
accurate and fast numerical simulations of these types of flows
require the development of efficient numerical solvers.

Since the prior work of \citet{bell1992second},
several authors proposed a second-order fractional time-step technique
for solving incompressible flows with large density variations
where Boussinesq approximation is no longer verified.
For instance,
\citet{almgren1998conservative} improve the convergence
by using adaptive mesh refinement and
\citet{calgaro2008hybrid} propose a hybrid method
where the mass conservation is solved by a finite volume method
and the velocity field is computed using a finite element method,
just to name a few.
The common feature of all the projection-like methods cited above
is that at each time-step the pressure
is determined by solving an elliptic equation
which can be written in its most general form as
\begin{equation}\label{eq:VDINSE.Lrho}%
  \opLrho(\phi) = \nabla \cdot \left(a \nabla \phi\right) = f
\end{equation}
due to the divergence-free constraint.
Especially,
$\phi$ is a scalar quantity related to the pressure,
$a = 1 / \rho$
with $\rho$ the approximation of the density for a given time
and $f$ some right hand-side term that changes with time.
\citet{concus1973use} first discuss of
how to tackle the numerical solution of \cref{eq:VDINSE.Lrho}
with an iterative algorithm.
For that purpose,
the authors suggest solving iteratively
the discrete counterpart of \cref{eq:VDINSE.Lrho}
where the system is rewritten as a Helmholtz-like equation
using change of variable.
The technique introduces a suitable choice of parameters
to accelerate the convergence.
In particular,
the authors show that for a second-order finite difference scheme,
the rate of convergence is independent of the mesh size
for a smooth $a(x, t)$ function.
However,
the choice of an optimal set of parameters
for a rapid convergence of \cref{eq:VDINSE.Lrho}
is directly connected to the condition number of the resulting operators.
The high convergence rate is therefore not guaranteed for high-order schemes.
Within the context of variable-density flows,
\citet{duffy2002improved},
\citet{cook2004mixing},
and more recently \citet{elouafa2021monolithic},
show that the discrete system associated with \cref{eq:VDINSE.Lrho}
is indeed ill-conditioned, which has for consequence
of significantly slow down the performance of widely used iterative solvers.
To overcome this difficulty,
\citet{guermond2009splitting} develop a numerical scheme
where \cref{eq:VDINSE.Lrho} is replaced by a Poisson equation
using a penalty function to verify the incompressibility constraint.
However,
the proposed method requires the introduction of an additional term
proportional to the divergence of the velocity fields
onto the mass equation and its amplitude has to be fixed.
More recently,
\citet{cook2005tera} develop
a hybrid Fourier spectral high-order compact finite-difference scheme
to investigate variable-density flows
which achieves tera-scalable computations
on massively parallel machines~\cite{cook2005tera}.
Using a conservative variable formulation,
they solve a Poisson equation for the pressure
by introducing an estimation of the update velocity.
As a consequence,
their projection onto the divergence-free vector space is no longer exact.
Finally, several authors suggest the use of preconditioning.
As underlined by \citet{elouafa2021monolithic},
the incomplete LU (ILU) preconditioning is generally used.
Nevertheless,
this preconditioner does not scale well in parallel implementation
with distributed memory.
In an effort to enhance convergence properties,
\citet{duffy2002improved} propose to combine
a multigrid technique with a preconditioning.
While the method improves the classical multigrid projection
used for example by \citet{ravier2004direct},
the efficiency of the preconditioning
is not discussed within the context of spectral methods.

The problem of the variable-coefficient discrete Poisson equation
is also considered by \citet{knikker2011comparative} for low-Mach-number flows
for which an extension of the projection-type method
proposed by \citet{bell1992second} is used.
Focusing on high-order finite difference schemes,
extensive numerical experiments are then carried out
by \citet{knikker2011comparative} to illustrate the performance of
various algorithms such as Conjugate Gradient (CG)-like methods
with different preconditioning techniques.
Besides,
the problem of ill-conditioned matrices associated
with spectral discretization of the Helmholtz equation
is discussed by \citet{haldenwang1984chebyshev}.
The authors present an iterative algorithm well-suited
or Chebyshev polynomial approximation
which uses a preconditioning built
on a second-order difference discretization of the Laplacian operator.
The more general case corresponding to \cref{eq:VDINSE.Lrho}
is not considered in the work of \citet{haldenwang1984chebyshev}.

From the above discussion, it appears that
the evaluation of an efficient preconditioner
combined with widely used iterative linear solvers for the resolution of
the variable-density incompressible Navier--Stokes equations (VDINSE)
with spectral spatial accuracy and second-order time-accurate scheme
has not yet been fully addressed.
Motivated by this question,
it is the objective of the present work to present a preconditioning technique
and its performance along with some iterative solvers
for solving the VDINSE system.

This paper presents a numerical method for solving VDINSE
that includes the motion induced by Fick's mass diffusion law,
with spectral spatial accuracy,
using a semi-implicit method for the viscous and diffusive term
and a second-order fractional time step to hold incompressibility.
It extends the previous work of \citet{dipierro2013projection}
where the influence of preconditioning is not addressed.
In a first section,
after having briefly presented the system of equations and numerical schemes,
the preconditioning technique based on the constant-density operator
is introduced for both the velocity and pressure equations.
In the second section,
the cost of the preconditioner is investigated through
numerical test cases carried out on the implicit systems
associated with both the velocity and pressure equations
for various iterative solvers.
The third section highlights the robustness of the methods
by time-marching the VDINSE system for representative numerical flow cases.

\section{\label{sec:equations}Governing equations and numerical schemes}

\subsection{Mathematical model}

We consider hereafter the motion of a viscous fluid in an inhomogeneous medium
that takes place in a bounded rectangular domain $\Omega$
(with boundaries noted $\partial\Omega$)
where the Cartesian coordinate system is defined by the $x, y, z$ axes
and in a time interval $t \in [0, T]$.
The mathematical model for
the variable-density incompressible Navier--Stokes equations (VDINSE)
used in the present contribution is detailed by
\citet{frank1955diffusion},
\citet{kazhikhov1977correctness},
\citet{antontsev1990boundary}
and \citet{guillen2007approximation}.
As in previous cited studies,
we introduce the mean density $\rho(\vec{x}, t)$
and the mean-volume velocity $\vec{u} = (u,v,w)\trans(\vec{x}, t)$,
then the dimensionless equations of motion read
\begin{subequations}\label{eq:VDINSE}%
  \begin{align}
    \label{eq:VDINSE.momentum}%
    \pdv{\vec{u}}{t} + \ugrad{\vec{u}} + \frac{\nabla{}p}{\rho}
      &= \vec{\zeta}(\rho, \vec{u}) + \vec{f}
    ,
    \\
    \label{eq:VDINSE.mass}%
    \odv{\rho}{t} + \ugrad{\rho}
      &= \frac{1}{\Reyn\Schm} \nabla^2 \rho
    ,
    \\
    \label{eq:VDINSE.divu}%
    \nabla \cdot \vec{u} &= 0
    \vphantom{\frac{1}{\rho\Reyn}}
    ,
    \\
    \label{eq:VDINSE.zeta}%
    \vec{\zeta}(\rho, \vec{u})
      &= \frac{1}{\rho\Reyn}\nabla^2\vec{u}
      + \frac{1}{\rho\Reyn\Schm}\bigr(\ugrad{\nabla\rho}
      + \left(\nabla\rho\cdot\nabla\right)\vec{u}\bigr)
    .
    \end{align}
\end{subequations}
Here, the mass diffusion is modeled according to the Fick's diffusion law.
In the momentum \cref{eq:VDINSE.zeta},
$p$ is a potential function analogous to the pressure
and $\vec{f}$ represents an external body force.
In \cref{eq:VDINSE.momentum} and \cref{eq:VDINSE.mass},
$\Reyn$ and $\Schm$ are the Reynolds and Schmidt numbers, respectively,
defined as
\begin{equation}
  \Reyn = \frac{\bar{\rho} U L}{\mu}
  ,
  \qquad
  \Schm = \frac{\mu}{\bar{\rho}\lambda}
  ,
\end{equation}
with $\mu$ and $\lambda$
the dynamic viscosity and mass diffusivity of the fluid respectively,
and $\bar\rho$, $U$, $L$
being characteristic density, velocity and length scales.

The system (\labelcref{eq:VDINSE}) is then closed with boundary conditions.
Two different types of boundary conditions are used in the present study:
no-slip and periodic boundary conditions.
For wall-bounded flows,
the same boundary conditions as in~\citet{guillen2007approximation} are
imposed:
\begin{equation}
  \vec{u}_{\vert\partial\Omega} = \mathbf{0}
  ,
  \qquad
  \frac{\partial\rho}{\partial\vec{n}}_{\vert\partial\Omega} = 0
  .
\end{equation}
Finally,
one may note that while many fundamental results
have been found with periodic boundary conditions%
~\cite{brachet1983small,kerr1985higher,vincent1991spatial},
such boundary conditions could also be used for spatially developing flows
by introducing a fringe region method~\cite{schlatter2005windowing}.

The system \labelcref{eq:VDINSE}
is solved using a fractional-step scheme with second-order accuracy.
The viscous and diffusive terms are discretized by
using a semi-implicit Crank--Nicholson scheme
(by extension of the prior study of \citet{bell1992second}):
\begin{subequations}\label{eq:VDINSE.num}%
  \begin{align}
    \label{eq:VDINSE.num.momentum}%
    \frac{\vec{u}^{n+1} - \vec{u}^n}{{\updelta}t} &=
      - \big[\ugrad{\vec{u}}\big]^{n+1/2}
      - \left[\frac{{\nabla}p}{\rho}\right]^{n+1/2}
      \\\nonumber&\quad\,
      + \frac12 \left(\vec{\zeta}(\rho^{n+1/2}, \vec{u}^{n+1})
      + \vec{\zeta}(\rho^{n+1/2}, \vec{u}^n)\right)
      ,
      \\
    \label{eq:VDINSE.num.mass}
    \frac{\rho^{n+1} - \rho^n}{{\updelta}t} &=
      - \big[\ugrad{\rho}\big]^{n+1/2}
      + \frac12 \frac1{\Reyn\Schm} \left(
          \nabla^2(\rho^{n+1})
        + \nabla^2(\rho^{n})
        \right)
    ,
    \\
    \label{eq:VDINSE.num.divu}
    \nabla \cdot \vec{u}^{n+1} &= 0
    \vphantom{-\left[\frac{\nabla p}{\rho}\right]^{n+1/2}}
    ,
  \end{align}
\end{subequations}
where ${\updelta}t$ is the time step,
the $n$ superscript denotes the solution at time $t^n=n{\updelta}t$
and $n+1/2$ represents a second-order approximation at time $t^{n+1/2}$.
Here, $[\ugrad{\vec{u}}]^{n+1/2}$ and $[\ugrad{\rho}]^{n+1/2}$
are estimated through an Adams--Bashforth scheme
whereas the viscous/mass diffusion term $\vec{\zeta}$ is computed with
$\rho^{n+1/2} = (\rho^{n+1} + \rho^n) / 2$
for stability considerations~\cite{tadjeran2007stability}.

Following \citet{bell1992second},
\cref{eq:VDINSE.num.momentum,eq:VDINSE.num.mass} are time-integrated by
introducing an intermediate velocity $\vecuAst$:
\begin{subequations}
  \begin{align}
    \label{eq:VDINSE.num.momentum.proj}%
    \frac{\vecuAst - \vec{u}^n}{{\updelta}t} &=
      - \big[\ugrad{\vec{u}}\big]^{n+1/2}
      - \frac{{\nabla}p^{n-1/2}}{\rho^{n+1/2}}
      + \frac12 \left(
          \vec{\zeta}(\rho^{n+1/2}, \vecuAst)
        + \vec{\zeta}(\rho^{n+1/2}, \vec{u}^n)
        \right)
    ,
    \\
    \label{eq:VDINSE.num.mass.proj}%
    \frac{\rho^{n+1}-\rho^n}{{\updelta}t} &= - \big[\ugrad{\rho}\big]^{n+1/2}
      + \frac12 \frac1{\Reyn\Schm} \left(
          \nabla^2(\rho^{n+1})+\nabla^2(\rho^{n})
        \right)
    .
  \end{align}
\end{subequations}
The velocity $\vec{u}^{n+1}$ is then updated
by performing the projection of $\vecuAst$ onto a divergence-free subspace
\begin{equation}\label{eq:VDINSE.num.proj}%
  \vec{u}^{n+1}
    = \vecuAst - {\updelta}t \frac{\nabla\phi}{\rho^{n+1/2}}
    = \opPrho(\vecuAst)
  ,
\end{equation}
where the scalar $\phi$ is solution of the equation
\begin{equation}
  \label{eq:VDINSE.Lrho.num}
  \opLrho(\phi)
    = \nabla \cdot \left(\frac{1}{\rho^{n+1/2}} \nabla \phi \right)
    = \frac{1}{{\updelta}t}\nabla \cdot \vecuAst
\end{equation}
with the same boundary condition used for the density field
(periodic or $(\partial\phi/\partial\vec{n})_{\vert\partial\Omega} = 0$)
to preserve precision order.

At this point,
the choice of the update pressure $p^{n+1/2}$
is crucial to truly verify a second-order accurate time scheme
as shown by \citet{brown2001accurate} in the case of constant-density flows.
By injecting $\vecuAst$
from \cref{eq:VDINSE.num.proj} into \cref{eq:VDINSE.num.momentum.proj}
and by comparing with the requested scheme from \cref{eq:VDINSE.num.momentum}
one can get the update pressure $p^{n+1/2}$ in gradient form
\begin{equation}\label{eq:VDINSE.num.gradp.update}%
  {\nabla}p^{n+1/2} = {\nabla}p^{n-1/2}
    + {\nabla}\phi
    - \frac{{\updelta}t}2 \rho^{n+1/2}
      \vec{\zeta}(\rho^{n+1/2}, {\nabla}\phi/\rho^{n+1/2})
\end{equation}
or in (equivalent) scalar form
\begin{equation}\label{eq:VDINSE.num.p.update}
  p^{n+1/2} =  p^{n-1/2} + \phi
    - (\nabla^2)^{-1} \Bigg(
        \nabla \cdot \bigg(
          \frac{{\updelta}t}2 \rho^{n+1/2} \vec{\zeta}(\rho^{n+1/2},
            \nabla\phi/\rho^{n+1/2})
        \bigg)
      \Bigg)
  ,
\end{equation}
with $\left(\nabla^2\right)^{-1}$ the inverse of Laplacian operator.
As mentioned by \citet{dipierro2013projection},
the projection operator $\opPrho$ is an exact projection operator.
Indeed,
\begin{equation}
  \opPrho(\vec{u})
    = \vec{u} - \frac{1}{\rho} \nabla\opLrho^{-1} \nabla \cdot \vec{u}
\end{equation}
and then,
for any divergence-free vector field $\vec{v}$
and any vector field $\vec{w}$ one gets:
\begin{equation}
  \opPrho(\vec{v}) = \vec{v}
  ,
  \qquad
  (\opPrho \circ \opPrho)(\vec{w}) = \opPrho(\vec{w})
  .
\end{equation}

\section{Elliptic solver}

The solution of an elliptic equation with constant coefficients
using spectral approximation
can be computed with fast and accurate algorithms
which exploit properties of the spectral basis%
~\cite{orszag1980spectral, haldenwang1984chebyshev, peyret2002spectral}.
The main difficulty for solving VDINSE
is that the pressure \cref{eq:VDINSE.Lrho.num}
is far more complicated than just a standard Poisson equation.
Especially,
such an elliptic problem discretized with spectral methods
yields to very ill-conditioned full matrices
for which direct solvers are inefficient.
As underlined by
\citet{canuto1995preconditioned} and \citet{peyret2002spectral},
systems such as \cref{eq:VDINSE.Lrho.num}
have to be solved with an iterative process.
Then,
the choice of an effective preconditioner
is crucial for increasing convergence speed
and improving the accuracy of iterative solvers~\cite{press1992numerical}.
Since the density variations considered here
are sufficiently smooth to be accurately projected onto a spectral basis,
this paper proposes to use
the inverse of the operator associated with constant-density flow
as preconditioner for three-dimensional Direct Numerical Simulation (DNS).
Within a Galerkin formulation,
the inverse matrix can be easily computed.
Indeed, constant-coefficients elliptic problems
lead to diagonal systems using Fourier decomposition
and to quasi-tridiagonal operators
with Chebyshev polynomials as basis functions.
Hence, for three-dimensional DNSs,
these elliptic problems are reduced to diagonal systems
with Fourier--Fourier--Fourier (F--F--F) decomposition
or $N^2$ quasi-tridiagonal systems
for Chebyshev--Fourier--Fourier (C--F--F) decomposition.
As presented
by \citet{dipierro2012spatiotemporal}and \citet{alizard2018space},
such decompositions are sufficient to study many fundamental configurations.
The case of Chebyshev--Chebyshev--Fourier
and fully Chebyshev decompositions are not treated here
because they lead to very large pentadiagonal or heptadiagonal matrices.
For the C--F--F case,
the solution of the preconditioned system
is determined by using the algorithm of \citet{thual1986transition}.
More specifically,
the solution of the tridiagonal system if computed
using the recurrence relation detailed in
\citet[appendix B]{peyret2002spectral}.
The algorithm requires $\Order(N)$ operations,
without storing the operator used for the preconditioning.
As mentioned by \cite{peyret2002spectral},
the algorithm is stable is the solution remains nicely bounded.
For that purpose,
sufficient conditions to ensure such a property
are given by \citet{peyret2002spectral},
which are satisfied for the following cases.
The efficiency of such preconditioning is discussed in the next section.

\subsection{Implicit viscous solver}

For illustration purposes,
we focus in this section on the one-dimensional case
where quantities are expressed using Chebyshev collocation discretization.
For that purpose,
the viscous linear operator associated
with the discrete \cref{eq:VDINSE.num.momentum.proj}
can be rewritten in the form
\begin{equation}\label{eq:VDINSE.Vrho}
  \opVrho(\vec{u}) = \left(
    \mat{I} - \frac{a}{\rho} \left(
        \mat{D}^2 + \frac{\partial^2 {\rho}}{\partial x^2}
      + \frac{\partial {\rho}}{\partial x} \mat{D}^1
    \right)
  \right) \vec{u}
\end{equation}
where $\mat{D}^k$ is the Chebyshev differentiation matrix of order $k$,
$\mat{I}$ the identity matrix,
and $\smash{a = \tfrac12 {\updelta}t / \Reyn}$
by assuming here $\Schm = 1$ for simplicity's sake.
We recall that an efficient resolution of $\opVrho(\vec{u}) = \vec{b}$
is strongly correlated with the condition number of the operator $\opVrho$.
We introduce the constant-density counterpart of $\opVrho$
\begin{equation}\label{eq:VDINSE.V}
  \opV = \left(\mat{I} - a \mat{D}^2 \right)
\end{equation}
which is used for preconditioning $\opVrho$.
For the numerical tests presented below, a classical top-hat like profile
\begin{equation}\label{eq:VDINSE.rho.hat}
  \rho(x) = 1 + \frac{s - 1}{2} \left(
    \tanh\left(\frac{x+x_0}{d}\right) - \tanh\left(\frac{x-x_0}{d}\right)
\right)
\end{equation}
is used for the density field
where $d = 0.1$ is the gradient length scale,
$s$ is the density ratio,
and $x_0$ is chosen as one fifth of the domain length.

\Cref{%
  fig:precond.visc.kappa.Cheb.vs_N,%
  fig:precond.visc.kappa.Cheb.vs_s,%
  fig:precond.visc.kappa.Cheb.vs_a}
show the condition number $\kappa$
--- computed by a singular values decomposition ---
of $\opVrho$ and $\opV$
when increasing the number of collocation points $N$,
the density ratio $s$,
and the viscous coefficient $a$.
The changes of $\kappa$ with respect to the variations in $N$, $s$ and $a$
for the left- and right-preconditioning
($\opV^{-1}\opVrho$ and $\opVrho\opV^{-1}$, respectively)
are also illustrated.
The same behavior is obtained with a Fourier expansion,
not detailed here for the sake of conciseness.
It is clear from the figures that even if $\kappa(\opVrho)$ and $\kappa(\opV)$
reach high values ($\Order(10^4)-\Order(10^8)$),
the proposed preconditioning technique greatly reduces the condition number.
Moreover,
those last two are quasi-independent of the collocation point number $N$
and remains small ($<\Order(10^2)$) even for larger values of $a$
(i.e.\ for viscous dominated flows) or $s$ (i.e.\ strongly stratified flows).

\def\figHeight{0.3\textheight}
\def\figWidth{0.7\textwidth}
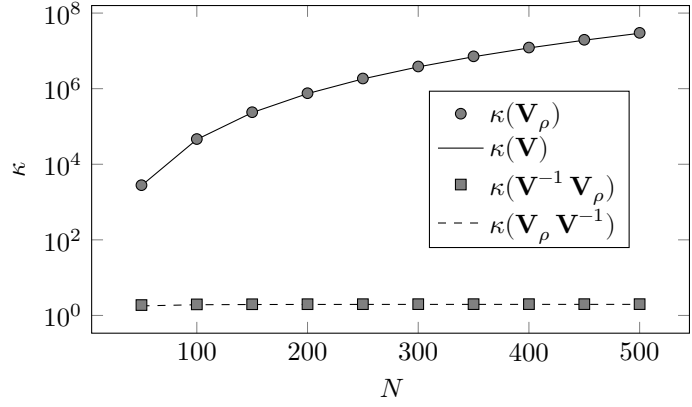
\begin{figure}[p]
  \centering
  \IfTikzLibraryLoaded{external}{
  \begin{tikzpicture}
    \pgfplotstableread[col sep=comma]
      {data/cond_visc_Cheb_vs_N_s0002_a1e-02.csv}\tableData
    \begin{semilogyaxis}[
      kappa axis, xlabel=$N$,
      height=\figHeight, width=\figWidth,
    ]
      \addplot table[x=n, y=kA] {\tableData};
      \addplot table[x=n, y=kP] {\tableData};
      \addplot table[x=n, y=klP]{\tableData};
      \addplot table[x=n, y=krP]{\tableData};
      \legend{%
        $\kappa(\opVrho)$,
        $\kappa(\opV)$,
        $\kappa (\opV^{-1} \opVrho)$,
        $\kappa (\opVrho \opV^{-1})$}
    \end{semilogyaxis}
  \end{tikzpicture}}{\includegraphics{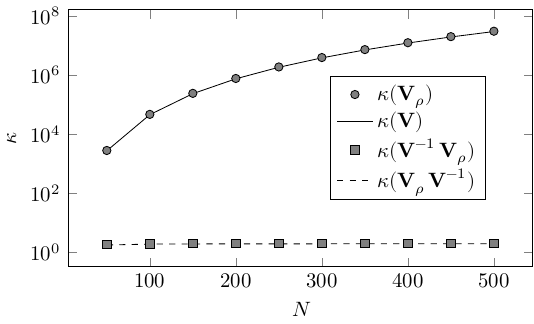}}
  \caption{\label{fig:precond.visc.kappa.Cheb.vs_N}%
    Condition number of the viscous operators
    versus the Chebyshev collocation points number $N$
    with $s = 2$ and $a = 0.01$.}
\end{figure}

\begin{figure}[p]
  \centering
  \IfTikzLibraryLoaded{external}{
  \begin{tikzpicture}
    \pgfplotstableread[col sep=comma]
        {data/cond_visc_Cheb_vs_s_N256_a1e-02.csv}\tableData
    \begin{loglogaxis}[
      kappa axis, xlabel=$s$,
      height=\figHeight, width=\figWidth,
    ]
      \addplot table[x=s, y=kA] {\tableData};
      \addplot table[x=s, y=kP] {\tableData};
      \addplot table[x=s, y=klP]{\tableData};
      \addplot table[x=s, y=krP]{\tableData};
      \legend{%
        $\kappa(\opVrho)$,
        $\kappa(\opV)$,
        $\kappa (\opV^{-1} \opVrho)$,
        $\kappa (\opVrho \opV^{-1})$}
    \end{loglogaxis}
  \end{tikzpicture}}{\includegraphics{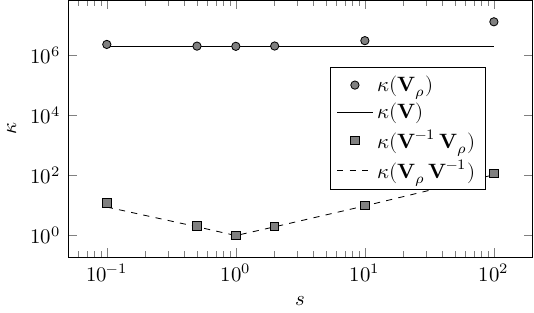}}
  \caption{\label{fig:precond.visc.kappa.Cheb.vs_s}%
    Condition number of the viscous operators
    versus the density ratio $s$
    with $a = 0.01$ and $N = 256$ Chebyshev collocation points.}
\end{figure}
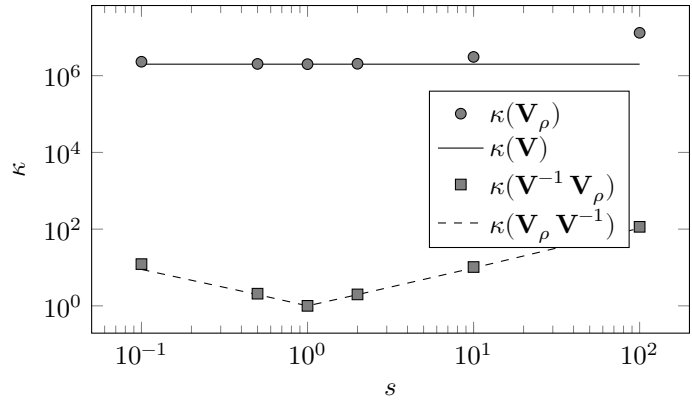

\begin{figure}[p]
  \centering
  \IfTikzLibraryLoaded{external}{
  \begin{tikzpicture}
    \pgfplotstableread[col sep=comma]
      {data/cond_visc_Cheb_vs_a_N256_s0002.csv}\tableData
    \begin{loglogaxis}[
      kappa axis, xlabel=$a$,
      height=\figHeight, width=\figWidth,
    ]
      \addplot table[x=a, y=kA] {\tableData};
      \addplot table[x=a, y=kP] {\tableData};
      \addplot table[x=a, y=klP]{\tableData};
      \addplot table[x=a, y=krP]{\tableData};
      \legend{%
        $\kappa(\opVrho)$,
        $\kappa(\opV)$,
        $\kappa (\opV^{-1} \opVrho)$,
        $\kappa (\opVrho \opV^{-1})$}
      \end{loglogaxis}
  \end{tikzpicture}}{\includegraphics{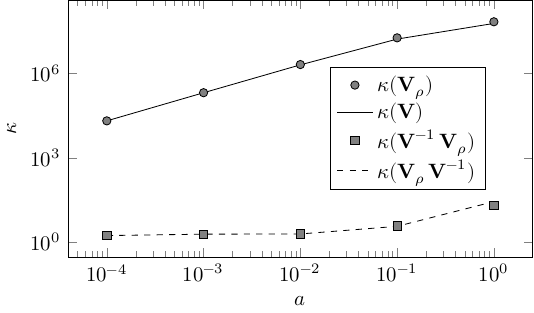}}
  \caption{\label{fig:precond.visc.kappa.Cheb.vs_a}%
    Condition number of the viscous operators
    versus $a$
    with $s=2$ and $N=256$ Chebyshev collocation points.}
\end{figure}
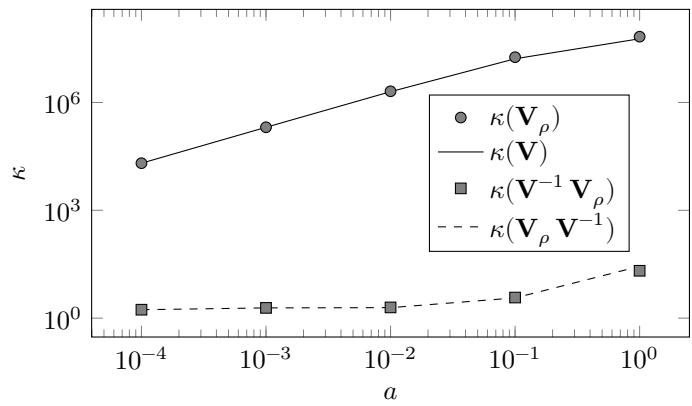

\begin{figure}[t]
  \centering
  \IfTikzLibraryLoaded{external}{
  \begin{tikzpicture}
    \begin{groupplot}[
      group style={
        group name=plot group,
        group size=2 by 1,
        ylabels at=edge left,
        horizontal sep=0.1\textwidth,
      },
      width=0.5\textwidth,
      ymode={log},
      conv axis,
    ]
      \pgfplotstableread[col sep=comma]
          {data/conv_visc_Four_N256_s0002_a1e-03.csv}\tableDataFour
      \pgfplotstableread[col sep=comma]
        {data/conv_visc_Cheb_N256_s0002_a1e-03.csv}\tableDataCheb
      \nextgroupplot[
        legend to name={fig:precond.visc.conv.legend},
      ]
        \addplot table[x=k_CG, y=r_CG]{\tableDataFour};
        \addplot table[x=k_BiCG, y=r_BiCG]{\tableDataFour};
        \addplot table[x=k_BiCGStab, y=r_BiCGStab]{\tableDataFour};
        \addplot table[x=k_RMR, y=r_RMR]{\tableDataFour};
        \addplot table[x=k_GMRES, y=r_GMRES]{\tableDataFour};
        \legend{CG, BiCG, BiCGStab, RMR, GMRES}
        \coordinate (c1) at (rel axis cs:0,1);
      \nextgroupplot
        \addplot table[x=k_CG, y=r_CG]{\tableDataCheb};
        \addplot table[x=k_BiCG, y=r_BiCG]{\tableDataCheb};
        \addplot table[x=k_BiCGStab, y=r_BiCGStab]{\tableDataCheb};
        \addplot table[x=k_RMR, y=r_RMR]{\tableDataCheb};
        \addplot table[x=k_GMRES, y=r_GMRES]{\tableDataCheb};
        \coordinate (c2) at (rel axis cs:1,1);
    \end{groupplot}
    \tikzset{subcaption/.style={
      text width=6cm,
      yshift=-0.05\textheight,
      align=center,
      anchor=north}}
    \node[subcaption] at (plot group c1r1.south)
      {\subcaption{Fourier differentiation}};
    \node[subcaption] at (plot group c2r1.south)
      {\subcaption{Chebyshev differentiation}};
    \coordinate (c3) at ($(c1)!.5!(c2)$);
    \node[above] at (c3 |- current bounding box.north)
      {\pgfplotslegendfromname{fig:precond.visc.conv.legend}};
  \end{tikzpicture}}{\includegraphics{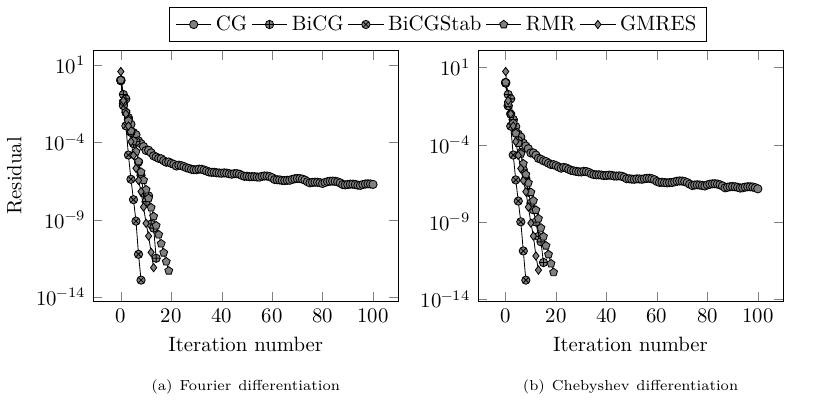}}
  \caption{\label{fig:precond.visc.conv}%
    Convergence history for the five preconditioned iterative solvers
    to solve $\opVrho(\vec{u}) = \vec{b}$
    with $a = 10^{-3}$ and $s = 2$
    with Fourier differentiation (left side)
    and Chebyshev differentiation (right side)
    and $N = 256$ collocation points.
    Convergence history for different parameters can be found
    in \labelcref{apdx:precond.visc.conv}.}
\end{figure}
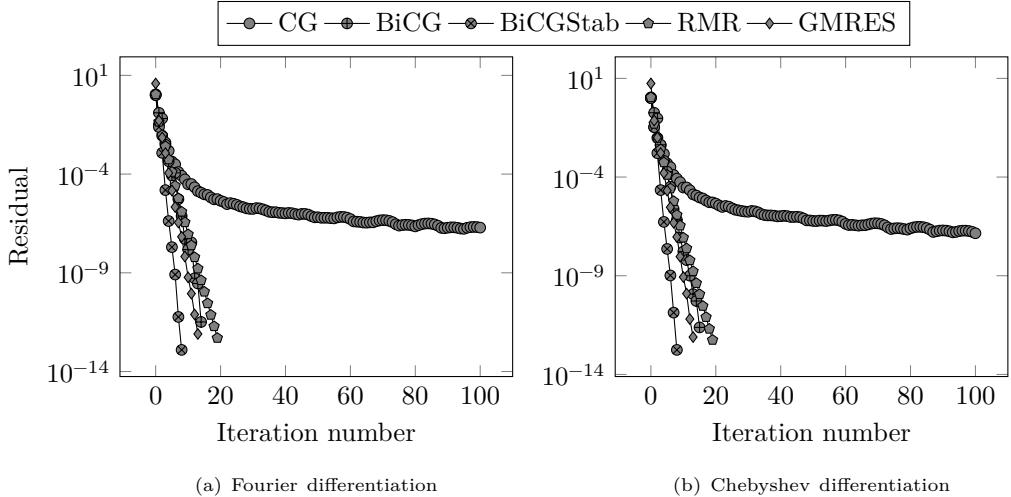

Let us now focus on the effect of such a preconditioning technique
onto the convergence speed of the algorithms
for the resolution of $\opVrho(\vec{u}) = \vec{b}$.
To this end, three iterative algorithms
--- based on the minimization of the residual
$\|\vec{b} - \opVrho(\vec{u})\|_2^2$ ---
are investigated:
\begin{itemize}
  \item Richardson Minimal Residual (RMR)~\cite{canuto1995preconditioned},
  \item Conjugate Gradient (CG)~\cite{press1992numerical},
  \item General Minimal Residual (GMRES)~\cite{saad1986gmres}.
\end{itemize}
One may note that the Biconjugate Gradient (BiCG)
and stabilized Biconjugate Gradient (BiCGStab) methods
are not proposed here, since for DNSs, matrices are never formed explicitly.
As a consequence,
computing the operator transposition
used in BiCG method yields some difficulties.
The convergence history of such algorithms
--- assuming a zero function as initial guess for any iterative process ---
is shown in \cref{fig:precond.visc.conv}
for $s = 2$ and $a = 10^{-2}$, as an example.
Similar investigations over a wide range of $(a, s)$
values are reported in \labelcref{apdx:precond.visc.conv}.
For these cases,
the convergence history of BiCG and BiCGStab are presented as well
for readers interested in other applications.
The independence of the convergence rate
with the number of collocation points $N$
has also been verified and is not detailed here for the sake of conciseness.

Results are summarized below.
The CG method provides the slower convergence
and do not reach a residual smaller than $10^{-6}$
(for all numerical tests performed here):
it turns out not to be competitive with other methods
in terms of accuracy and computational time.
This behavior is probably due to
the non-symmetric properties of the operator associated
with \cref{eq:VDINSE.Vrho}
For $s = 2$,
RMR and GMRES methods converge both towards
a residual value of $\approx 10^{-12}$
in about $20$ iterations
independently of $a$
(see \cref{fig:precond.visc.conv.vs_a} in \labelcref{apdx:precond.visc.conv}).
For $a = 10^{-2}$,
it can be seen in \cref{fig:precond.visc.conv.vs_s}
(see \labelcref{apdx:precond.visc.conv})
that the RMR does not converge when $s > 10$.
For this value of $a$,
the GMRES algorithm is the only method that
reaches a residual value smaller than $10^{-9}$
for extremely stratified flows ($s > 100$).
One may note that the convergence of the GMRES algorithm is here
sufficiently fast to not require a restart process.
Let $M$ the number of iterations necessary to reach a convergence criterion,
the complexity of RMR method is $\Order(N \times M)$ whereas the
GMRES complexity is $\Order(N \times M^2)$,
due to the additional orthonormalization process.
Hence, the latter is expected to be much slower even for a few iterations
(especially if the numerical method is parallelized
with a domain decomposition strategy).
This can be seen in \cref{tab:precond.visc.time}
which shows the computational time of RMR and GMRES methods
without an explicit construction of differentiation matrices.

\begin{table}[t]
  \centering
  \begin{tabular}{|c|c|c|}
    \hline
    Number of iterations & RMR & GMRES    \\\hline
     $5$ & $5.6~10^{-3}$ & $1.0~10^{-2}$  \\\hline
    $10$ & $1.1~10^{-2}$ & $3.4~10^{-2}$  \\\hline
    $20$ & $2.0~10^{-2}$ & $9.7~10^{-2}$  \\\hline
    $40$ & $4.2~10^{-2}$ & $2.2~10^{-1}$  \\\hline
  \end{tabular}
  \caption{\label{tab:precond.visc.time}%
    Computational time in seconds of the preconditioned RMR and GMRES
    to solve $\opVrho(\vec{u}) = \vec{b}$,
    with $s = 1000$ and $N = 256$ Chebyshev basis collocation points.
    Computations are performed with an Intel Xeon E5-2670 CPU
    mounted on a Dell PowerEdge C8220 Compute Node.}
\end{table}

Finally, to accelerate the convergence,
the previously converged solution $\vec{u}^n$
is used as an initial guess for solving $\vec{u}^{n+1}$.
This leads to a fast convergence for both the RMR and GMRES methods:
only $3$ to $5$ iterations
are required to obtain a residual value varying
from $\Order(10^{-12})$ to $\Order(10^{-14})$.
Hence,
the RMR method should be adopted
for solving \cref{eq:VDINSE.num.momentum.proj}
when the density ratio is lower than $s = 10$
and the GMRES method should be adopted
for steeper cases.

\subsection{Pressure equation}

As previously mentioned,
\cref{eq:VDINSE.Lrho.num} for the scalar $\phi$ is very ill-conditioned
and has to be solved accurately for ensuring mass conservation.
The scalar $\Phi = \phi / \sqrt\rho$
is introduced in order to
rewrite \cref{eq:VDINSE.Lrho.num} as a modified Helmholtz equation
\begin{equation}\label{eq:VDINSE.Krho}
  \opKrho(\Phi) = \nabla^2\Phi + \left(
    \frac{\nabla^2\rho}{2 \rho} - \frac{3}{4}\frac{||\nabla \rho||^2}{\rho^2}
  \right) \Phi = \sqrt{\rho} f
\end{equation}
with $f = \left(\nabla \cdot \vecuAst\right) / {\updelta}t$.
One may recall that \citet{concus1973use} use a similar technique
for the general case of elliptic equations with variable coefficients.
It's worth noting that when using fully periodic boundary conditions,
$\Phi$ remains periodic.
When wall boundary conditions
\begin{equation}
  \frac{\partial\rho}{\partial\vec{n}}_{|\partial\Omega} = 0,
  \qquad
  \frac{\partial\phi}{\partial\vec{n}}_{|\partial\Omega} = 0,
\end{equation}
are considered,
a Neumann condition is also obtained for $\Phi$ and reads
\begin{equation}
  \frac{\partial\Phi}{\partial\vec{n}}_{|\partial\Omega} = 0.
\end{equation}

As for the viscous operator,
the operator $\opKrho$ is solved iteratively and is preconditioned
by the inverse of the discrete Laplacian operator
$\smash{\left(\nabla^2\right)^{-1}}$
which is nothing else than the inverse of $\opKrho$ with constant density.
Following the same line of thought as in the previous section,
the behavior of the preconditioned operator is firstly characterized through
a one-dimensional test case discretized
with Chebyshev collocation matrices~\cite{peyret2002spectral}.

Using the density profile described by \cref{eq:VDINSE.rho.hat},
the condition numbers of these implicit operators (preconditioned or not)
are shown in
\cref{fig:precond.pres.kappa.Cheb.vs_N,fig:precond.pres.kappa.Cheb.vs_s}
versus either the collocation point number or the density ratio.
One may notice that $\opKrho$ and $\opLrho$ are both ill-conditioned
even if the use of $\opKrho$ instead of $\opLrho$ leads to some improvements.
The figures show that both left- and right-preconditioning
provide a condition number $\kappa = \Order(\textrm{$10^0$--$10^2$})$.
Especially,
the condition number is seen independent of the collocation point number $N$.
In \cref{fig:precond.pres.kappa.Cheb.vs_s},
one may also observe that
$\smash{\kappa(\left(\nabla^2\right)^{-1}\opKrho)
  < \kappa(\left(\nabla^2\right)^{-1}\opLrho)}$
for $0.1 \leq s \leq 100$.

The same algorithms as in the previous section (CG, RMR and GMRES) are studied
and compared with the fixed point (FP) algorithm
proposed by \citet{dipierro2013projection}.
The convergence history of these iterative solvers is shown
in \cref{fig:precond.pres.conv} for $s = 2$ and $a = 10^{-2}$
and in \labelcref{apdx:precond.pres.conv}
for different values of the density ratio $s$.
The independence of the convergence rate
with the number of collocation points $N$
has been verified and is not detailed here for the sake of conciseness.
Results are now summarized.
As for the viscous operator,
the CG method is inaccurate in comparison to other methods,
especially, the residual exhibits a plateau around $0.5$
for all test cases.
The GMRES has the fastest convergence,
leading to a decay of the residual
by 12 decades in about 10 iterations for all test cases.
Besides,
the RMR needs 4 times more iterations
in order to reach the same level of convergence
as the one obtained using the GMRES for $s = 2$
and does not converge for higher density ratios.
Finally,
the FP method~\cite{dipierro2013projection} converges faster
than the RMR method for $s = 2$
and has the same convergence rate as the GMRES for more stratified cases
while being less accurate (its residual saturates around $\Order(10^{-9})$).
In that regard,
the RMR method should be adopted for solving \cref{eq:VDINSE.Krho}
when the density ratio is around $s = 2$
while the GMRES should be used for higher density ratios.

\def\figHeight{0.29\textheight}
\def\figWidth{0.7\textwidth}
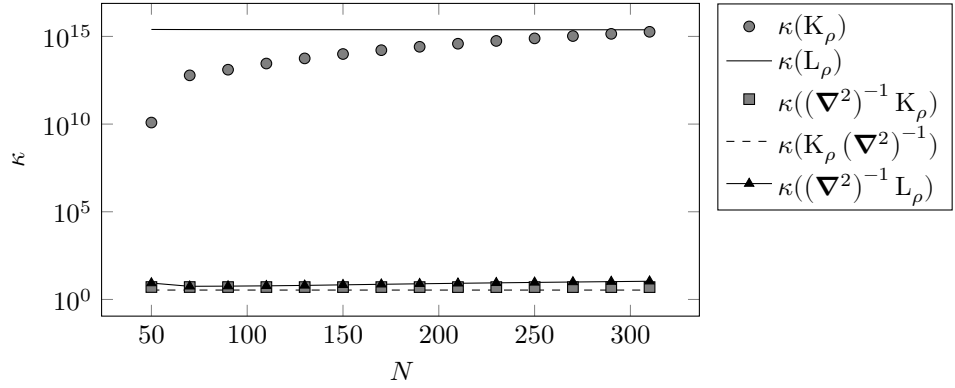
\begin{figure}[p]
  \centering
  \IfTikzLibraryLoaded{external}{
  \begin{tikzpicture}
    \pgfplotstableread[col sep=comma]
      {data/cond_pres_Cheb_vs_N_s0002.csv}\tableData
    \begin{semilogyaxis}[
      kappa axis, xlabel=$N$,
      width=\figWidth, height=\figHeight,
      legend pos=outer north east,
    ]
      \addplot table[x=n, y=kK]{\tableData};
      \addplot table[x=n, y=kL]{\tableData};
      \addplot table[x=n, y=klPK]{\tableData};
      \addplot table[x=n, y=krPK]{\tableData};
      \addplot table[x=n, y=klPL]{\tableData};
      \legend{%
        $\kappa(\opKrho)$,
        $\kappa(\opLrho)$,
        $\kappa(\left(\nabla^2\right)^{-1}\opKrho)$,
        $\kappa(\opKrho\left(\nabla^2\right)^{-1})$,
        $\kappa(\left(\nabla^2\right)^{-1}\opLrho)$,
      }
    \end{semilogyaxis}
  \end{tikzpicture}}{\includegraphics{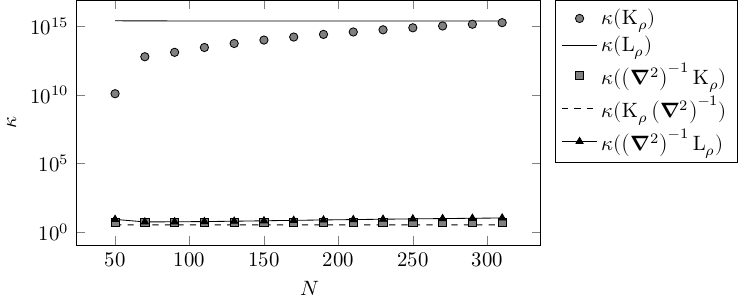}}
  \caption{\label{fig:precond.pres.kappa.Cheb.vs_N}%
    Condition number of the pressure operators
    versus Chebyshev collocation point number $N$
    with $s = 2$.}
\end{figure}

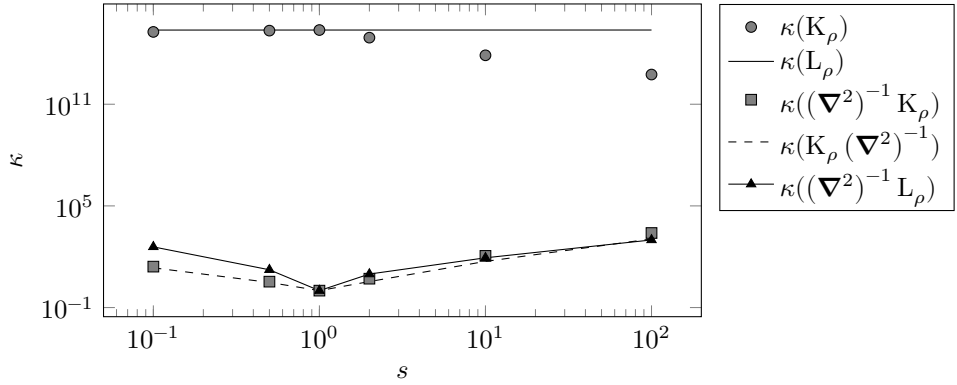
\begin{figure}[p]
  \centering
  \IfTikzLibraryLoaded{external}{
  \begin{tikzpicture}
    \pgfplotstableread[col sep=comma]
      {data/cond_pres_Cheb_vs_s_N256.csv}\tableData
    \begin{loglogaxis}[
      kappa axis, xlabel=$s$,
      width=\figWidth, height=\figHeight,
      legend pos=outer north east,
    ]
    \addplot table[x=s, y=kK]{\tableData};
    \addplot table[x=s, y=kL]{\tableData};
    \addplot table[x=s, y=klPK]{\tableData};
    \addplot table[x=s, y=krPK]{\tableData};
    \addplot table[x=s, y=klPL]{\tableData};
    \legend{%
      $\kappa(\opKrho)$,
      $\kappa(\opLrho)$,
      $\kappa(\left(\nabla^2\right)^{-1}\opKrho)$,
      $\kappa(\opKrho\left(\nabla^2\right)^{-1})$,
      $\kappa(\left(\nabla^2\right)^{-1}\opLrho)$,
    }
    \end{loglogaxis}
  \end{tikzpicture}}{\includegraphics{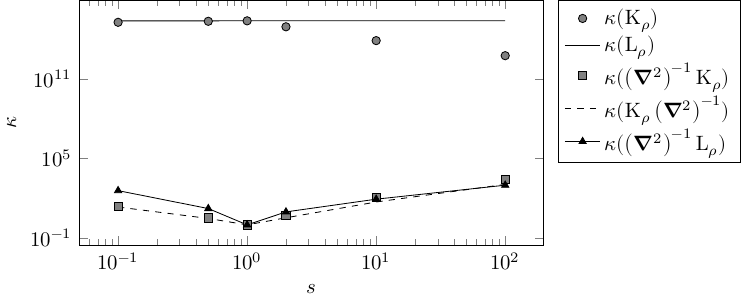}}
  \caption{\label{fig:precond.pres.kappa.Cheb.vs_s}%
    Condition number of the pressure operators
    versus density ratio $s$
    with $N = 256$ Chebyshev collocation points.}
\end{figure}

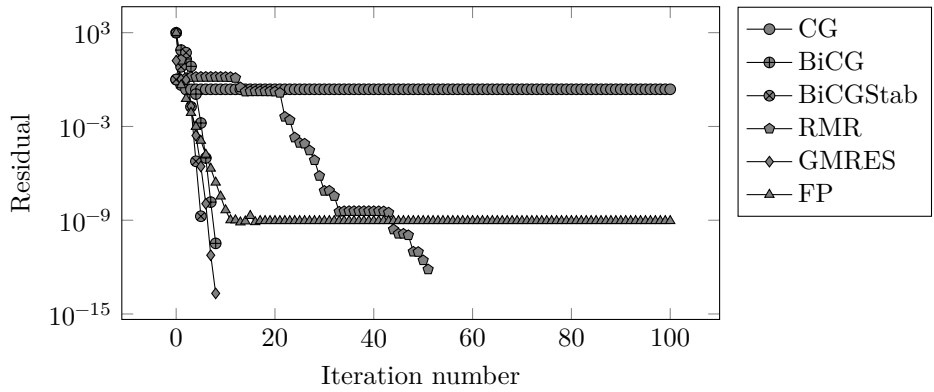
\begin{figure}[p]
  \centering
  \IfTikzLibraryLoaded{external}{
  \begin{tikzpicture}
    \pgfplotstableread[col sep=comma]
      {data/conv_pres_Cheb_N256_s0002.csv}\tableData
    \begin{semilogyaxis}[
      conv axis,
      legend pos=outer north east,
      legend columns={1},
      legend cell align=left,
      width=\figWidth, height=\figHeight,
    ]
      \addplot table[x=k_CG, y=r_CG]{\tableData};
      \addplot table[x=k_BiCG, y=r_BiCG]{\tableData};
      \addplot table[x=k_BiCGStab, y=r_BiCGStab]{\tableData};
      \addplot table[x=k_RMR, y=r_RMR]{\tableData};
      \addplot table[x=k_GMRES, y=r_GMRES]{\tableData};
      \addplot table[x=k_FP, y=r_FP]{\tableData};
      \legend{CG, BiCG, BiCGStab, RMR, GMRES, FP}
    \end{semilogyaxis}
  \end{tikzpicture}}{\includegraphics{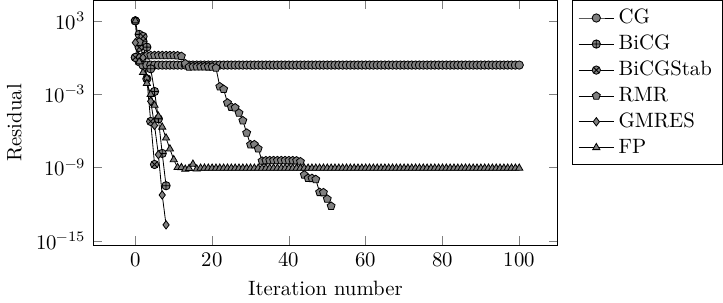}}
  \caption{\label{fig:precond.pres.conv}%
    Convergence history of
    the different preconditioned iterative solvers
    to solve the modified pressure \cref{eq:VDINSE.Krho}
    with $s = 2$, and $N = 256$ Chebyshev collocation points.
    Convergence history for different parameters can be found
    in \labelcref{apdx:precond.pres.conv}.}
\end{figure}

\section{\label{sec:results}Results and validation}

\subsection{Time accuracy and mass conservation}

The derivation of an exact solution of the VDINSE system
is not straightforward.
To the best of the authors' knowledge, none has been found until now.
Hence,
to estimate the accuracy of the presented method,
mass source and body-force terms are added to the VDINSE equations such that
\begin{subequations}\label{eq:VDINSE.solexact}%
  \begin{align}
    u_   \text{e}(x, y, z, t) &= \cos(t)\sin(x)\cos(y)\cos(z)    \\
    v_   \text{e}(x, y, z, t) &= \cos(t)\cos(x)\sin(y)\cos(z)    \\
    w_   \text{e}(x, y, z, t) &= -2\cos(t)\cos(x)\cos(y)\sin(z)  \\
    p_   \text{e}(x, y, z, t) &= \sin(t)\sin(2x)\sin(2y)\sin(2z) \\
    \rho_\text{e}(x, y, z, t) &= 1
      + \tfrac{s - 1}{2}\left(1 + \sin(t)\sin(x)\sin(y)\sin(z)\right)
  \end{align}
\end{subequations}
defines an exact solution of the modified VDINSE,
where $s$ is (an arbitrary positive) density ratio.
One may note that this corresponds to a Taylor--Green vortex
with an imposed pressure and density field.

Let ($\vec{u}(T)$, $p(T)$, $\rho(T)$)
be the numerical solution extracted at a time $t = T$
computed by marching forward in time the
\cref{%
  eq:VDINSE.num.momentum.proj,%
  eq:VDINSE.num.mass.proj,%
  eq:VDINSE.num.proj,%
  eq:VDINSE.num.p.update}
with a time step ${\updelta}t$
and an initial condition matching the system \labelcref{eq:VDINSE.solexact}
for $t = 0$.
Then,
the temporal errors at $t = T$
are defined with respect to the exact solution as
$\epsilon_{\vec{u}} = ||\vec{u}(T) - \vec{u}_\text{e}(T)||_2$,
$\epsilon_p = ||p(T) - p_\text{e}(T)||_2$
and $\epsilon_\rho = ||\rho(T) - \rho_\text{e}(T)||_2$.
The corresponding errors are shown in \cref{fig:VDINSE.err}
for $128\times128\times128$ grid discretization points
and with $\Reyn = 1$, $\Schm = 1$, $s = 10$ and $T = \pi / 4$.
\Cref{fig:VDINSE.err} shows that the numerical solution is,
as expected, globally second-order accurate in time.
Additionally,
the spectral accuracy has been verified by running the same simulation
with different grid sizes.
Machine precision ($10^{-14}$) is reached from a $32^3$ mesh
for the divergence of the velocity field.

From a more physical point of view,
the total mass of the considered system
$ M = \int_\Omega \rho \,\odif{\Omega}$
has to be strictly conserved throughout the simulation.
From Fick's law and \cref{eq:VDINSE.mass},
the mass evolution over time is directly given
by the divergence of velocity field
\begin{equation}
  \odv{M}{t} = \int_\Omega \nabla \cdot \vec{u} \,\odif{\Omega}.
\end{equation}
\Cref{fig:VDINSE.divu} shows the time evolution
of the $L^2$ norm of $\nabla\cdot\vec{u}$
for the previous test case.
One can observe that
this quantity is nearly equal to the machine precision
all along the simulation.

\def\figHeight{0.45\textheight}
\def\figWidth{0.7\textwidth}
\begin{figure}[p]
  \centering
  \IfTikzLibraryLoaded{external}{
  \begin{tikzpicture}
    \pgfplotstableread{data/VDINSE_err.dat}\tableData
    \begin{axis}[
      legend style={
        at={(0.90, 0.25)},
        anchor=east,
      },
      grid=both,
      xmode=log,
      ymode=log,
      xlabel={${\updelta}t$},
      ylabel={Error},
      width=\figWidth,
      height=\figHeight,
    ]
      \addplot [dashed]
        table[x=dt, y=dt2]{\tableData};
      \addplot [every mark/.append style={fill=gray}, mark=*]
        table[x=dt, y=err_u]{\tableData};
      \addplot [every mark/.append style={fill=gray}, mark=triangle*]
        table[x=dt, y=err_p]{\tableData};
      \addplot [mark=asterisk]
        table[x=dt, y=err_rho]{\tableData};
      \legend{%
        $\updelta t^2$,
        $\epsilon_u$,
        $\epsilon_p$,
        $\epsilon_\rho$,
      }
    \end{axis}
  \end{tikzpicture}}{\includegraphics{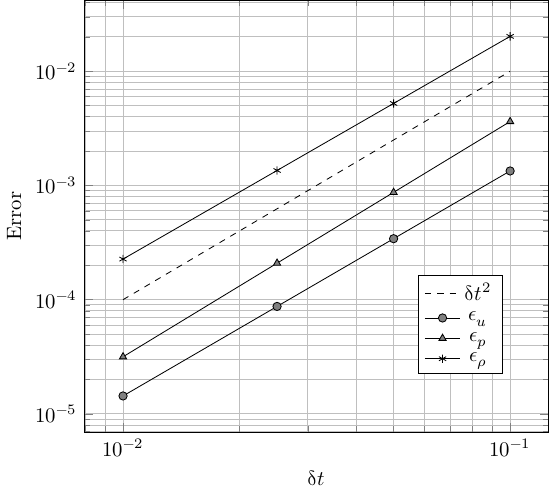}}
  \caption{\label{fig:VDINSE.err}%
    Temporal errors at $t = \pi/4$
    with respect to the exact solution of the modified VDINSE
    \labelcref{eq:VDINSE.solexact}
    versus the time-step ${\updelta}t$
    with $\Reyn = 1$, $\Schm = 1$, $s = 10$
    and $N = 128^3$ discretization grid points.}
\end{figure}
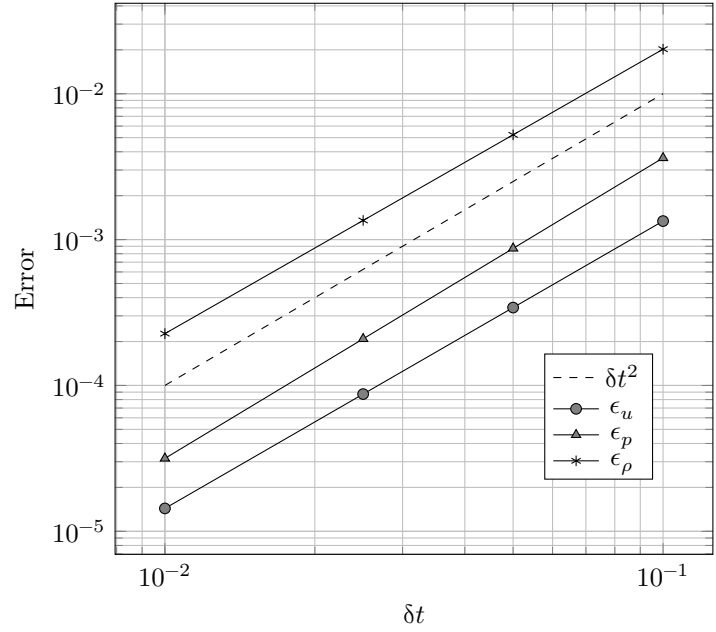

\begin{figure}[p]
  \centering
  \IfTikzLibraryLoaded{external}{
  \begin{tikzpicture}
    \begin{axis}[
      ylabel={$\|\nabla\cdot\vec{u}\|_2$},
      xlabel={$t$},
      width=\figWidth,
      height=\figHeight,
    ]
      \addplot [black] table {data/VDINSE_divu.dat};
    \end{axis}
  \end{tikzpicture}}{\includegraphics{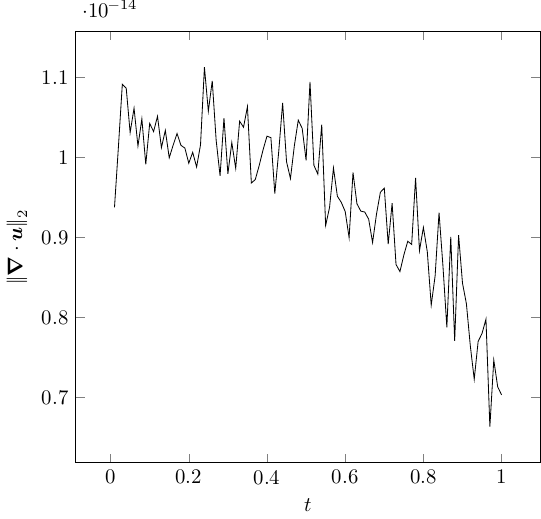}}
  \caption{\label{fig:VDINSE.divu}%
    $L_2$ norm of the divergence of the velocity field
    versus time
    during a simulation of the modified VDINSE
    \cref{eq:VDINSE.solexact}
    with $\Reyn = 1$, $\Schm = 1$, $s = 10$, ${\updelta}t = 0.01$
    and $N = 128^3$ discretization grid points.}
\end{figure}
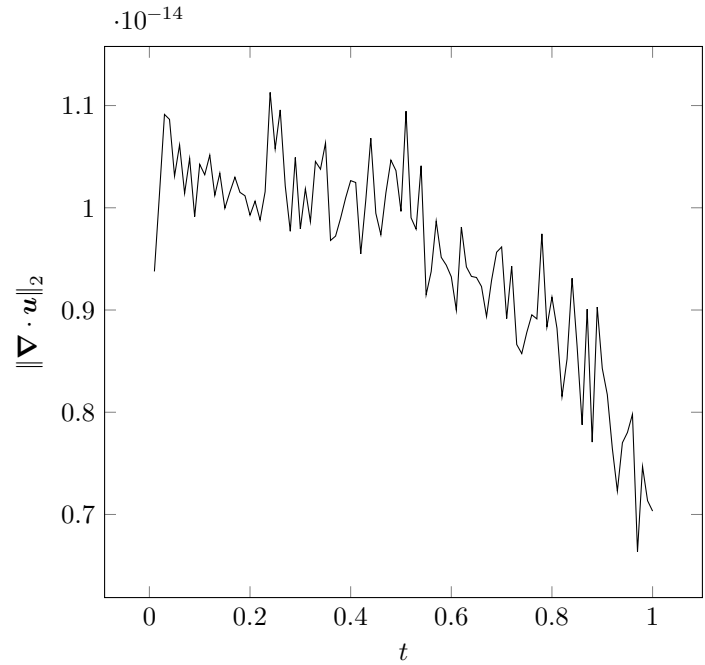

\subsection{3D solver performance}

The performance of the presented method
is now discussed by running three-dimensional DNSs
associated with the Taylor--Green vortex discussed above
(see system \labelcref{eq:VDINSE.solexact}).
The computational times for the pressure iterative solver
are shown in \cref{tab:VDINSE.pres.time}
for the different solvers investigated in the previous section
(FP, RMR and GMRES)
and with $128\times128\times128$ discretization points.
Unlike the 1D case detailed in the previous section,
a Fourier (periodic) decomposition is used.
For the preconditioning,
a singular value decomposition is performed to compute
a pseudo-inverse of the discrete Laplacian.
It is equivalent to remove the first spectral coefficient
associated with the polynomial of degree one.
It imposes that
$\phi$ and $\Phi$ are zero-mean.

One can see from \cref{tab:VDINSE.pres.time}
that RMR and GMRES solvers are faster than the FP method.
This is due to the problem formulation:
$\opKrho$ needs only two physical/spectral transforms
whereas four are needed to evaluate $\opLrho$.
To accelerate the convergence,
the solution $\smash{\phi^{n-1/2}}$ is used as an initial condition
for evaluating $\phi$ at the time-step $n+1/2$.
In agreement with the condition number of the preconditioned operators,
the performance of each solver is also independent of the density ratio $s$.
Hence, it
validates the performance of the method even for large density ratios.
Moreover,
the computational time per iteration per thread and per cell
is in the order of magnitude of the CPUs clock time.
Hereafter,
the GMRES algorithm is adopted
for the DNS test cases for illustration purposes.

\begin{table}[t]
  \centering
  \begin{tabular}{|l|c|c|c|}\hline
    Iterative solver & FP (\cite{dipierro2013projection}) & RMR & GMRES
    \\\hline
    \multicolumn{4}{|c|}{$s = 2$} \\\hline
    Iterations & 7 & 5 & 6 \\\hline
    Time & 0.5973 & 0.2124 & 0.2480 \\\hline
    Time/iteration & 0.0853 & 0.0425 & 0.0413 \\\hline
    Time/iteration/thread & 0.0107 & 0.0053 & 0.0052 \\\hline
    Time/iteration/thread/cells & 5.1e-09 & 2.5e-09 & 2.5e-09 \\\hline
    \multicolumn{4}{|c|}{$s = 6$} \\\hline
    Iterations & 9 & 7 & 8 \\\hline
    Time & 0.7737 & 0.3045 & 0.3573 \\\hline
    Time/iteration & 0.0860 & 0.0435 & 0.0447 \\\hline
    Time/iteration/thread & 0.0107 & 0.0054 & 0.0056 \\\hline
    Time/iteration/thread/cell & 5.1e-09 & 2.6e-09 & 2.7e-09\\\hline
    \multicolumn{4}{|c|}{$s = 10$} \\\hline
    Iterations & 10 & 8 & 8 \\\hline
    Time & 0.8232 & 0.3477 & 0.3547 \\\hline
    Time/iteration & 0.0823 & 0.0435 & 0.0443 \\\hline
    Time/iteration/thread & 0.0103 & 0.0054 & 0.0055 \\\hline
    Time/iteration/thread/cells & 4.9e-09 & 2.6e-09 & 2.6e-09 \\\hline
  \end{tabular}
  \caption{\label{tab:VDINSE.pres.time}%
    Computational time in seconds for iterative solvers
    for the modified pressure \cref{eq:VDINSE.Krho}
    and for the test case flow (system \labelcref{eq:VDINSE.solexact})
    Different values of the density ratio $s$ are investigated,
    with $128 \times 128 \times 128$ discretization points.
    Computations are performed on the 8 CPU cores of an
    Intel Xeon E5-2670 mounted on a Dell PowerEdge C8220 Compute Node.
    CPU Times are averaged over 10 time steps
    ($10{\updelta}t \approx \pi / 2$).}
\end{table}

\subsection{Rayleigh--Taylor instability}

\begin{figure}
  \centering
  \IfTikzLibraryLoaded{external}{
  \begin{tikzpicture}
    \begin{groupplot}[
      group style={
        group name=plot group,
        group size=3 by 2,
        horizontal sep=0.05\textwidth,
        vertical sep=0.15\textwidth,
      },
      width=0.45\textwidth,
      tick style={draw=none},
      ytick=\empty,
      xtick=\empty,
    ]
      \nextgroupplot[
        enlargelimits = false,
        axis on top,
        axis equal image,
      ]
        \addplot graphics[xmin=-1, xmax=1, ymin=-1, ymax=1]
          {figures/RTI2D_1mode_s10_t3.0.pdf};
      \nextgroupplot[
        enlargelimits = false,
        axis on top,
        axis equal image,
      ]
        \addplot graphics[xmin=-1, xmax=1, ymin=-1, ymax=1]
          {figures/RTI2D_1mode_s10_t4.0.pdf};
      \nextgroupplot[
        enlargelimits = false,
        axis on top,
        axis equal image,
      ]
        \addplot graphics[xmin=-1, xmax=1, ymin=-1, ymax=1]
          {figures/RTI2D_1mode_s10_t5.0.pdf};
        \coordinate (c2) at (rel axis cs:1,1);
      \nextgroupplot[
        enlargelimits = false,
        axis on top,
        axis equal image,
      ]
        \addplot graphics[xmin=-1, xmax=1, ymin=-1, ymax=1]
          {figures/RTI2D_multimode_s10_t1.0.pdf};
      \nextgroupplot[
        enlargelimits = false,
        axis on top,
        axis equal image,
      ]
        \addplot graphics[xmin=-1, xmax=1, ymin=-1, ymax=1]
          {figures/RTI2D_multimode_s10_t2.0.pdf};
      \nextgroupplot[
        enlargelimits = false,
        axis on top,
        axis equal image,
      ]
        \addplot graphics[xmin=-1, xmax=1, ymin=-1, ymax=1]
          {figures/RTI2D_multimode_s10_t3.0.pdf};
    \end{groupplot}
    \node[above, yshift=0.02\textwidth] at (plot group c2r1.north)
      {(a) Single-mode perturbation};
    \node[above, yshift=0.02\textwidth] at (plot group c2r2.north)
      {(b) Multi-mode perturbation};
    \tikzset{subcaption/.style={
      text width=2cm,
      yshift=-0.01\textheight,
      anchor=north,}
    }
    \node[subcaption] at (plot group c1r1.south) {(a1) $t = 3$};
    \node[subcaption] at (plot group c2r1.south) {(a2) $t = 4$};
    \node[subcaption] at (plot group c3r1.south) {(a3) $t = 5$};
    \node[subcaption] at (plot group c1r2.south) {(b1) $t = 1$};
    \node[subcaption] at (plot group c2r2.south) {(b2) $t = 2$};
    \node[subcaption] at (plot group c3r2.south) {(b3) $t = 3$};
  \end{tikzpicture}}{\includegraphics{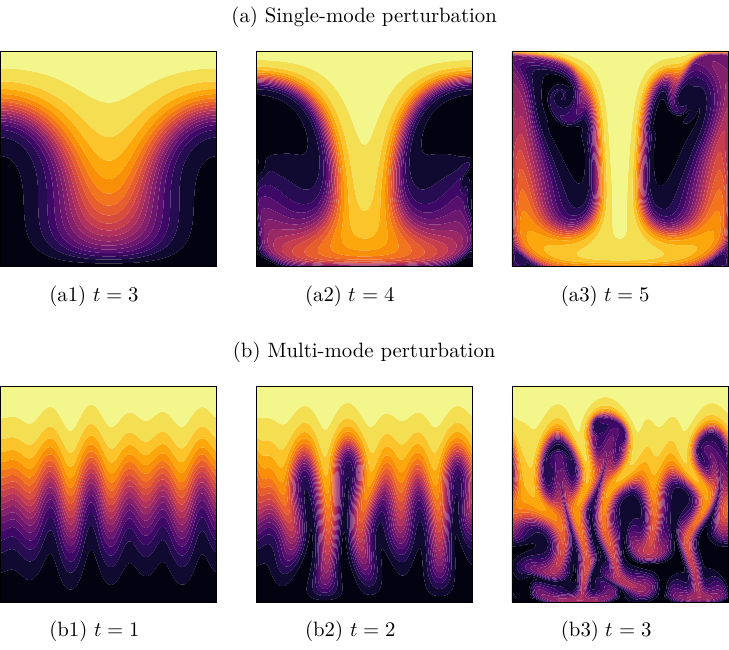}}
  \caption{\label{fig:RTI}%
    Contour plot of the density field at different times
    from 2D RTI simulations
    using the preconditioned RMR for both the viscous and pressure operators
    with $s = 10$, $\Reyn = 1000$, $\Frou = 1$, $\Schm = 1$
    and $256^2$ grid points.
    In~(a1), (a2), and (a3)
    a single-mode perturbation is injected in the initial density field.
    In~(b1), (b2), and (b3)
    a multi-mode perturbation is injected in the initial density field.
    \Cref{fig:RTI.conv} presents the convergence history
    for the pressure operator corresponding to~(b3).}
\end{figure}

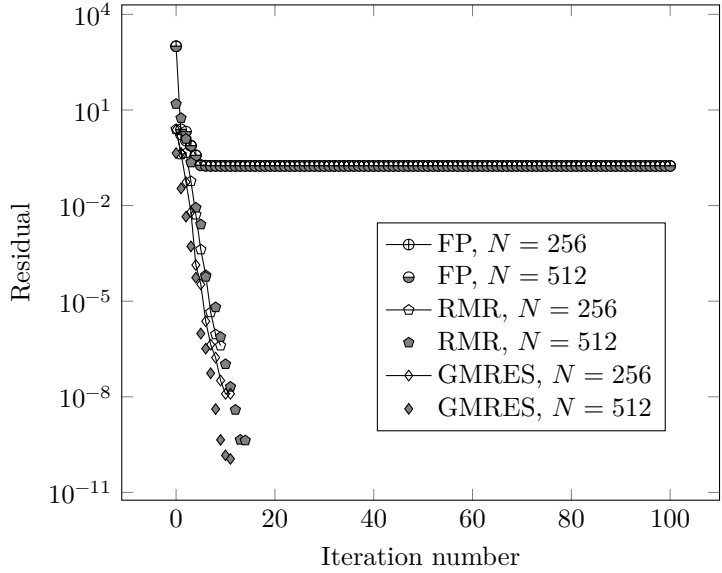
\begin{figure}
  \centering
  \IfTikzLibraryLoaded{external}{
  \begin{tikzpicture}
    \pgfplotstableread[col sep=comma]
      {data/RTI2D_multimode_s10_t3.0_conv.dat}\tableData
    \begin{semilogyaxis}[
      conv axis,
      legend cell align=left,
      legend style={at={(0.91, 0.35)}, anchor=east, legend columns=1},
      cycle list={
        mark=oplus\\
        only marks, mark color=gray, mark=halfcircle\\
        mark=pentagon\\
        only marks, every mark/.append style={fill=gray}, mark=pentagon*\\
        mark=diamond\\
        only marks, every mark/.append style={fill=gray}, mark=diamond*\\
      },
      width=0.7\textwidth,
    ]
      \foreach \solv in {FP, RMR, GMRES} {
        \foreach \n in {256, 512} {
          \addplot table[x=k_\solv_\n, y=r_\solv_\n]{\tableData};
            \addlegendentryexpanded{\solv, $N = \n$}
        }
      }
    \end{semilogyaxis}
  \end{tikzpicture}}{\includegraphics{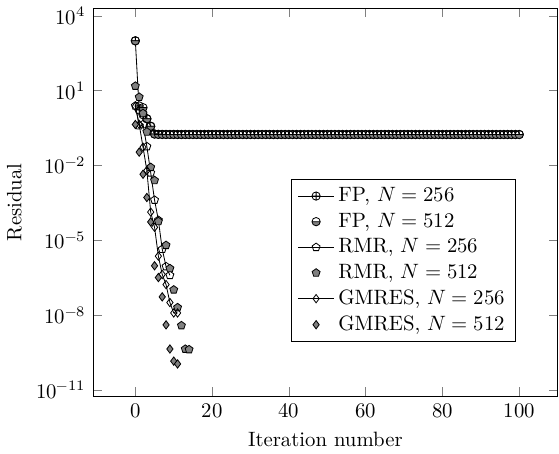}}
  \caption{\label{fig:RTI.conv}%
    Convergence history of the different preconditioned iterative solvers
    to solve the modified pressure equation \cref{eq:VDINSE.Krho}
    in the case~(b3) of \cref{fig:RTI}.}
\end{figure}

First, the effectiveness and robustness of the present method is illustrated
through a classical configuration of a 2D Rayleigh--Taylor instability (RTI):
a heaver fluid is maintained above a lighter one
and gravity acceleration is in the opposite direction of the density gradient.
No-slip boundary conditions are used on the top and bottom walls,
while periodic boundary conditions are used on the left and right boundaries.
The initial density profile is a smoothed Heavyside function
in the vertical direction
disturbed horizontally with a small amplitude perturbation $\eta$
\begin{equation}
  \rho(x,z,t=0)
    = 1 + \frac{s - 1}{2} \left( \tanh\left(\frac{z}{\delta}
      + \eta(x, L_x)\right) + 1 \right)
  .
\end{equation}
The computational box size is fixed to $[2L_x, 2L_z] = [2, 2]$
(discretized with $N^2$ grid points varying from $256^2$ to $512^2$).
The initial thickness is $\delta = 0.2$
and $s$ is the density ratio.
The initial velocity field is set to zero.
Gravity is added through a volumetric force
\begin{equation}
  \vec{f} = - \Frou^{-1} \vec{e}_z
\end{equation}
with $\Frou$ the Froude number.
Within this configuration,
computations without preconditioning do not converge,
and numerical oscillations take over in a few iterations.
\Cref{fig:RTI}
shows density contours at different times for both a single-mode%
~((a1), (a2), (a3))
\begin{equation}
  \eta = 0.1 \cos\left(\pi \frac{x}{L_x} \right)
\end{equation}
and a multi-mode initial perturbation~((b1), (b2), (b3))
\begin{equation}
  \eta = 0.1 \left(
        \sin\left(13 \pi \frac{x}{L_x} \right)
      + \cos\left(17 \pi \frac{x}{L_x} \right)
      + \cos\left(19 \pi \frac{x}{L_x} \right)
    \right)
\end{equation}
with dimensionless numbers $\Reyn=1000$, $\Frou=1$, $\Schm=1$ and $s = 10$.
In the single-mode case,
one can see the characteristic mushroom of the Rayleigh--Taylor instability
growing downward in the middle of the computational domain.
As the heavier fluid penetrates the lighter fluid,
the interface rolls up into two vortices
as presented in \citet{bell1992second}.
In the multi-mode case,
the waves begin to grow independently of one another
before they start to interact strongly with each other
reproducing patterns similar to those presented in \citet{bell1992second}.
In both cases,
it can be seen that the interface is clearly defined
without numerical oscillations,
which validates the robustness of the method.
\Cref{fig:RTI.conv} presents the convergence history
of the three studied methods (FP, RMR and GMRES)
when dealing with the modified pressure \cref{eq:VDINSE.Krho}
for the same range of parameters
and a multiple-mode initial perturbation at time $t = 3.0$.
It can be seen that the convergence rate is independent of the mesh size
and both the RMR and GMRES algorithms converge towards a residual value
lower than $10^{-7}$
which demonstrate the effectiveness of the method
in dealing with steep problems.
For this case, the FP method does not achieve a good convergence.

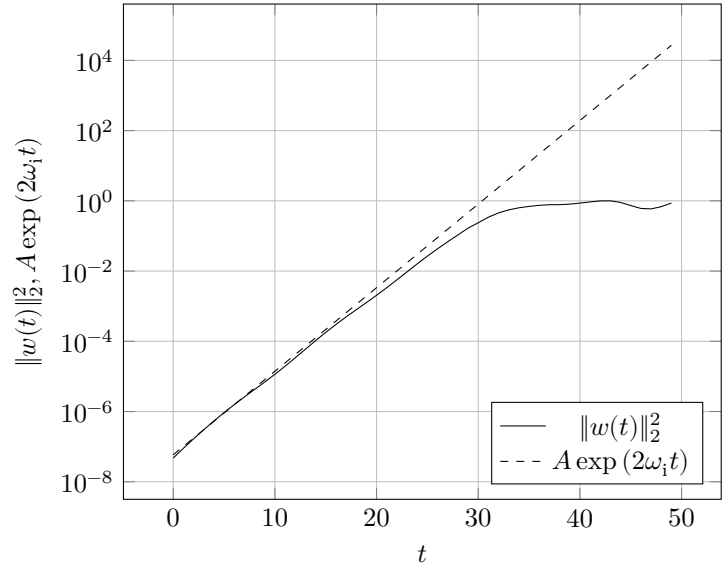
\begin{figure}[p]
  \centering
  \IfTikzLibraryLoaded{external}{
  \begin{tikzpicture}
    \pgfplotstableread{data/VDSwirl_wp2.dat}\tableData
    \begin{axis}[
      grid=both,
      no markers,
      ymode=log,
      ylabel={$\|w(t)\|^2_2, A\exp\left(2\omega_\text{i}t\right)$},
      xlabel=$t$,
      legend pos=south east,
      width=0.7\textwidth,
    ]
      \addplot [black]
        table {\tableData};
      \addplot [black, dashed]
        table [x index=0, y index=2] {\tableData};
      \legend{%
        $\|w(t)\|^2_2$,
        $A\exp\left(2\omega_\text{i}t\right)$
      };
    \end{axis}
  \end{tikzpicture}}{\includegraphics{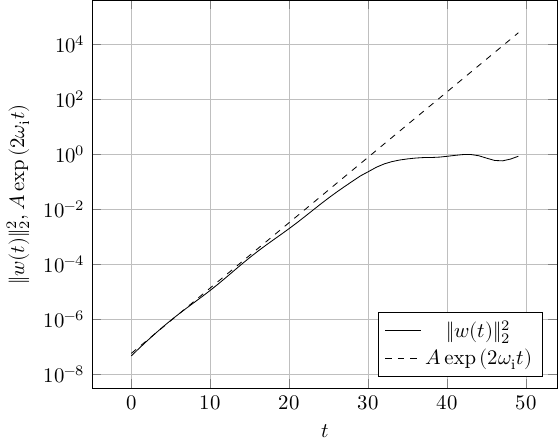}}
  \caption{\label{fig:Swirl.growthRate}%
    Evolution of the norm of the axial velocity perturbation
    during a DNS of a dense swirling jet
    with $\Reyn = 100$, $\Schm = 1$, $s = 2$, $q = 0.5$, $m = -4$ and $k = 1$.
    The growth rate obtained is $\omega_\text{i,DNS} = 0.275$, the growth rate
    predicted by the linear instability analysis
    is $\omega_\text{i} = 0.274$.}
\end{figure}

\begin{figure}[p]
  \centering
  \resizebox{!}{0.7\textwidth}{
  \IfTikzLibraryLoaded{external}{
  \begin{tikzpicture}[rotate around y=-30, rotate around x=0]
    \node[anchor=south east, inner sep=0] at (0, 0)
      {\includegraphics[width=\textwidth]{figures/VDSwirl_isosurf.png}};
    \draw[->] (-3, 1, 0) -- (-2, 1, 0) node[above] {$x$};
    \draw[->] (-3, 1, 0) -- (-3, 2, 0) node[right] {$z$};
    \draw[->] (-3, 1, 0) -- (-3, 1, 1) node[above] {$y$};
  \end{tikzpicture}}{\includegraphics{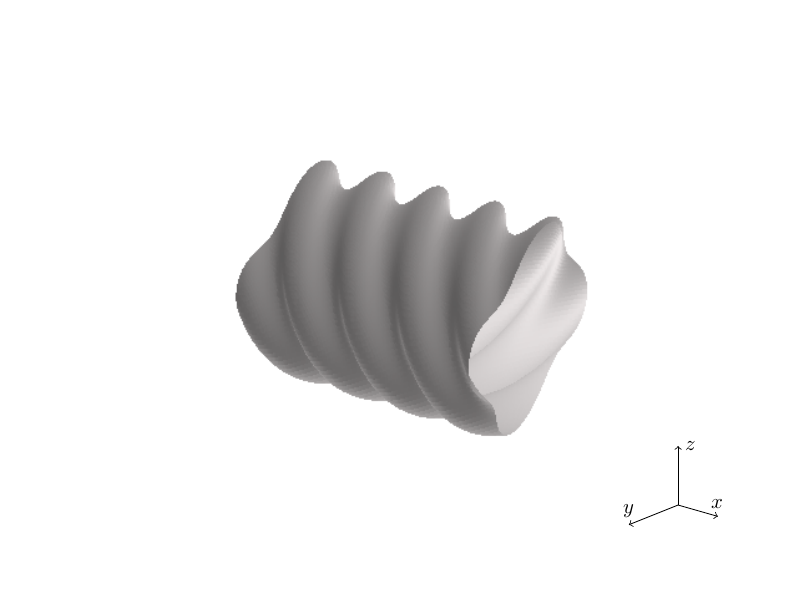}}}
  \caption{\label{fig:Swirl.isosurf}%
    Isosurfaces of the density field perturbation
    in the non-linear saturated regime at $t = 40$
    from a DNS of a dense swirling jet
    with $\Reyn = 100$, $\Schm = 1$, $s = 2$, $q = 0.5$, $m = -4$ and $k = 1$.}
\end{figure}

The reliability of the proposed method is now illustrated
through a representative three-dimensional variable-density flow case.
In that respect,
the variable-density swirling jet studied by \citet{dipierro2013projection}
is selected.
The Reynolds number --- based on the base flow stream-wise velocity ---
is fixed to $\Reyn = 100$ and
the others dimensionless control parameters are set to
$\Schm = 1$, $s = 2$ and $q = 0.5$.
The domain size is $2\pi \times 2\pi \times 4\pi$.
We consider the following equilibrium state:
\begin{subequations}\label{eq:Jet.swirl}%
  \begin{align}
    \rho_{0}(r) &= 1 + (s - 1) \exp{-r^2}
    ,
    \\
    \omega_x(r) &= 2 q \exp{-r^2}
    ,
    \\
    V_x(r) &= \exp{-r^2},
  \end{align}
\end{subequations}
where $\vec{V}_0 = (0, V_\theta, V_x)\trans$ is the base velocity field
(with$V_\theta\vec{e}_\theta =
  - \left(\nabla^2\right)^{-1}\left(\nabla \times \omega_x\vec{e}_x\right)$),
$\omega_x$ is the axial vorticity,
and $q$ is the swirl number.
We then capture the dynamics of a perturbation
$\phi(r)\exp{(\imath(kx + m\theta))}$
superimposed to the base flow with a very small amplitude.
In that purpose, $256^3$ discretization points are used.
It is found that the dynamics is driven
for long times by the azimuthal mode $m=4$
and the fundamental stream-wise wave number $k=1$
in agreement with the linear stability theory
(LST, see~\cite{dipierro2013projection}
for details about the linear stability analysis).
\Cref{fig:Swirl.growthRate}
shows the evolution of the axial velocity perturbation.
We then compare the exponential growth rate
extracted from the DNS database with the linear stability results.
In that case,
the linear stability analysis gives a temporal growth rate
$\omega_\text{i} = 0.274$.
For long times, we found a temporal growth rate from the DNS
\begin{equation}
  \omega_\text{i,DNS}
    = \frac1{2 (t_1 - t_0)}
        \log{\left(\frac{\|w(t_1)\|^2_2}{\|w(t_0)\|^2_2}\right)}
    = 0.2746 \approx 0.275
\end{equation}
which nearly matches the one computed from the LST.
The latter is calculated from the norm of the axial velocity perturbation $w$
with $t_0 = 2.0$ and $t_1 = 5.0$.
\Cref{fig:Swirl.isosurf} shows the isosurfaces of the density field
in the non-linear saturated regime associated with such flow case.
The interface is well-defined and no oscillations appear.
It further illustrates the robustness of the present method to deal with
representative variable-density flow case.

\section{Conclusions}

Performances of preconditioners
based on the constant-density operators
for the numerical resolution of
the variable-density and incompressible Navier--Stokes equations
with spectral accuracy and second-order time accuracy are investigated.
We first observed that test cases presented here did not converge
without any preconditioning,
which motivated this study.
It is shown that these preconditioners
highly reduce the condition number of
the variable-density elliptic operators
for both the pressure and velocity fields equations.
Coupled with widely used iterative solvers
(Conjugate Gradient, Richardson Minimal Residual,
and  General Minimal Residual),
we give strong evidence that
the proposed numerical methods highly enhance the convergence
of the implicit systems.
Finally,
the precision and the robustness of the method is further illustrated
on some representative
variable-density flow cases.

We thus believe that the present study can serve as a guide
for the development of faster and more accurate DNS incompressible solvers
that take into account large density variations.
For the extension of our method
to high-performance simulation on distributed-memory computers,
one may recall that the major drawback of spectral methods
lies in their global approximation.
However,
some solutions are suggested in the literature
and successfully numerically tested.
For instance,
one may cite the work of some members of the same team dealing with
high-scalability spectral code on high-performance distributed-memory%
~\cite{montagnier2012towards}.
Especially,
the authors show that the parallelization strategy for spectral schemes
based on a domain decomposition method
--- where the computational domain is subdivided
along spatial directions into subdomains ---
exhibits a good scalability and a very fast wall-clock time per iteration
on HPC platforms.
In the present work, the three-dimensional computations use the same strategy.

\section{Acknowledgements}

The authors thank the ``Fédération Lyonnaise de Modélisation et Sciences
Numérique'' for providing computational facilities on the computer center
P2CHPD at ``Université Claude Bernard Lyon 1''.

The authors also thank the group ``LyonCalcul'' for fruitful discussions.

\appendix

\section{\label{apdx:precond.visc.conv}%
  Viscous operator iterative solvers convergence history}

In this appendix,
we show the convergence history
of the CG, BiCG, BiCGStab, RMR and GMRES algorithms
for the resolution of the viscous equation $\opVrho(\vec{u}) = \vec{b}$,
where the viscous operator $\opVrho$ is defined \cref{eq:VDINSE.Vrho}
A wide range of $(a, s)$ is investigated.

\section{\label{apdx:precond.pres.conv}%
  Pressure operator iterative solvers convergence history}

In this appendix,
we show the convergence history
of the CG, BiCG, BiCGStab, RMR, GMRES and FP algorithms
for the resolution of the modified pressure \cref{eq:VDINSE.Krho}
A wide range of density ratios $s$ is investigated.

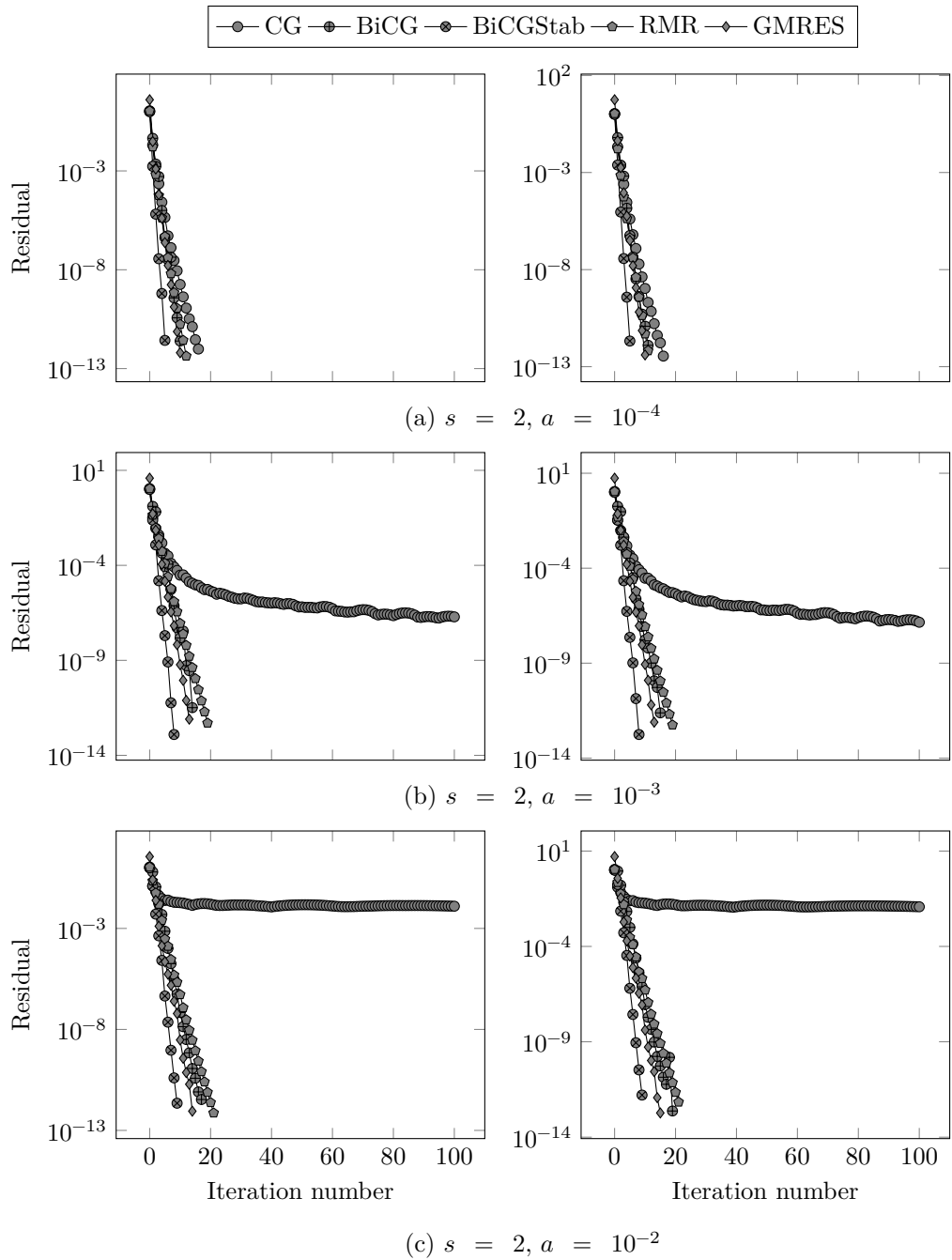
\begin{figure}[p]
  \centering
  \IfTikzLibraryLoaded{external}{
  \begin{tikzpicture}
    \begin{groupplot}[
      group style={
        group name=plot group,
        group size=2 by 3,
        ylabels at=edge left,
        xlabels at=edge bottom,
        x descriptions at=edge bottom,
        horizontal sep=0.1\textwidth,
      },
      width=0.5\textwidth,
      height=0.3\textheight,
      ymode={log},
      conv axis,
    ]
      \nextgroupplot[
        legend to name={fig:precond.visc.conv.vs_a.legend}
      ]
        \pgfplotstableread[col sep=comma]
          {data/conv_visc_Four_N256_s0002_a1e-04.csv}\tabledataI
        \addplot table[x=k_CG, y=r_CG]{\tabledataI};
        \addplot table[x=k_BiCG, y=r_BiCG]{\tabledataI};
        \addplot table[x=k_BiCGStab, y=r_BiCGStab]{\tabledataI};
        \addplot table[x=k_RMR, y=r_RMR]{\tabledataI};
        \addplot table[x=k_GMRES, y=r_GMRES]{\tabledataI};
        \legend{CG, BiCG, BiCGStab, RMR, GMRES}
        \coordinate (c1) at (rel axis cs:0,1);
      \nextgroupplot
        \pgfplotstableread[col sep=comma]
          {data/conv_visc_Cheb_N256_s0002_a1e-04.csv}\tabledataII
        \addplot table[x=k_CG, y=r_CG]{\tabledataII};
        \addplot table[x=k_BiCG, y=r_BiCG]{\tabledataII};
        \addplot table[x=k_BiCGStab, y=r_BiCGStab]{\tabledataII};
        \addplot table[x=k_RMR, y=r_RMR]{\tabledataII};
        \addplot table[x=k_GMRES, y=r_GMRES]{\tabledataII};
        \coordinate (c2) at (rel axis cs:1,1);
      \nextgroupplot
        \pgfplotstableread[col sep=comma]
          {data/conv_visc_Four_N256_s0002_a1e-03.csv}\tabledataIII;
        \addplot table[x=k_CG, y=r_CG]{\tabledataIII};
        \addplot table[x=k_BiCG, y=r_BiCG]{\tabledataIII};
        \addplot table[x=k_BiCGStab, y=r_BiCGStab]{\tabledataIII};
        \addplot table[x=k_RMR, y=r_RMR]{\tabledataIII};
        \addplot table[x=k_GMRES, y=r_GMRES]{\tabledataIII};
      \nextgroupplot
        \pgfplotstableread[col sep=comma]
          {data/conv_visc_Cheb_N256_s0002_a1e-03.csv}\tabledataIV;
        \addplot table[x=k_CG, y=r_CG]{\tabledataIV};
        \addplot table[x=k_BiCG, y=r_BiCG]{\tabledataIV};
        \addplot table[x=k_BiCGStab, y=r_BiCGStab]{\tabledataIV};
        \addplot table[x=k_RMR, y=r_RMR]{\tabledataIV};
        \addplot table[x=k_GMRES, y=r_GMRES]{\tabledataIV};
      \nextgroupplot
        \pgfplotstableread[col sep=comma]
          {data/conv_visc_Four_N256_s0002_a1e-02.csv}\tabledataV;
        \addplot table[x=k_CG, y=r_CG]{\tabledataV};
        \addplot table[x=k_BiCG, y=r_BiCG]{\tabledataV};
        \addplot table[x=k_BiCGStab, y=r_BiCGStab]{\tabledataV};
        \addplot table[x=k_RMR, y=r_RMR]{\tabledataV};
        \addplot table[x=k_GMRES, y=r_GMRES]{\tabledataV};
      \nextgroupplot
        \pgfplotstableread[col sep=comma]
          {data/conv_visc_Cheb_N256_s0002_a1e-02.csv}\tabledataVI;
        \addplot table[x=k_CG, y=r_CG]{\tabledataVI};
        \addplot table[x=k_BiCG, y=r_BiCG]{\tabledataVI};
        \addplot table[x=k_BiCGStab, y=r_BiCGStab]{\tabledataVI};
        \addplot table[x=k_RMR, y=r_RMR]{\tabledataVI};
        \addplot table[x=k_GMRES, y=r_GMRES]{\tabledataVI};
    \end{groupplot}
    \coordinate (c3) at ($(c1)!.5!(c2)$);
    \node[above] at (c3 |- current bounding box.north)
      {\pgfplotslegendfromname{fig:precond.visc.conv.vs_a.legend}};
    \tikzset{subcaption/.style={
      text width=6cm,
      yshift=-0.01\textheight,
      align=center,
      anchor=north}}
    \node[subcaption]
      at ($(plot group c1r1.south)!.5!(plot group c2r1.south)$)
      {(a) $s = 2,\, a = 10^{-4}$};
    \node[subcaption]
      at ($(plot group c1r2.south)!.5!(plot group c2r2.south)$)
      {(b) $s = 2,\, a = 10^{-3}$};
    \node[subcaption,yshift=-0.05\textheight]
      at ($(plot group c1r3.south)!.5!(plot group c2r3.south)$)
      {(c) $s = 2,\, a = 10^{-2}$};
  \end{tikzpicture}}{\includegraphics{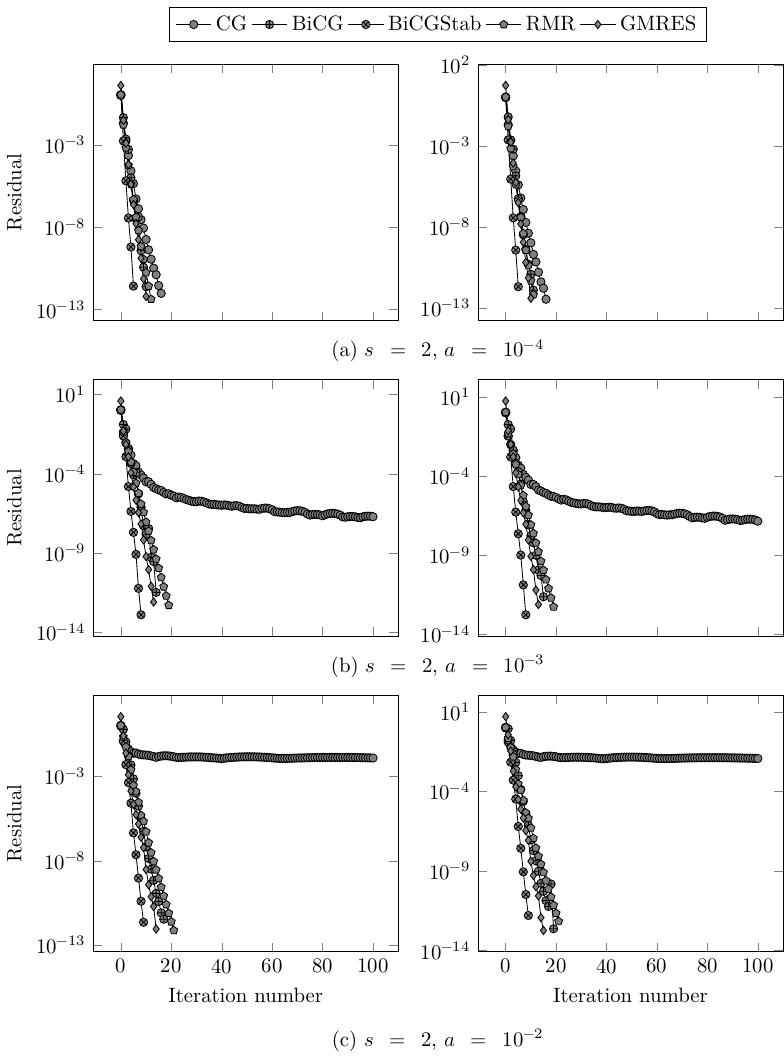}}
  \caption{\label{fig:precond.visc.conv.vs_a}%
    Convergence history for the five preconditioned iterative solvers
    to solve $\opVrho(\vec{u}) = \vec{b}$
    with $s = 2$ and different values of $a$,
    with Fourier differentiation (left side)
    and Chebyshev differentiation (right side)
    and $N = 256$ collocation points.}
\end{figure}

\begin{figure}[p]
  \centering
  \IfTikzLibraryLoaded{external}{
  \begin{tikzpicture}
    \begin{groupplot}[
      group style={
        group name=plot group,
        group size=2 by 3,
        ylabels at=edge left,
        xlabels at=edge bottom,
        x descriptions at=edge bottom,
        horizontal sep=0.1\textwidth,
      },
      width=0.5\textwidth,
      height=0.3\textheight,
      ymode={log},
      conv axis,
    ]
      \nextgroupplot[
        legend to name={fig:precond.visc.conv.vs_s.legend}
      ]
        \pgfplotstableread[col sep=comma]
          {data/conv_visc_Four_N256_s0010_a1e-02.csv}\tabledataI
        \addplot table[x=k_CG, y=r_CG]{\tabledataI};
        \addplot table[x=k_BiCG, y=r_BiCG]{\tabledataI};
        \addplot table[x=k_BiCGStab, y=r_BiCGStab]{\tabledataI};
        \addplot table[x=k_RMR, y=r_RMR]{\tabledataI};
        \addplot table[x=k_GMRES, y=r_GMRES]{\tabledataI};
        \legend{CG, BiCG, BiCGStab, RMR, GMRES}
        \coordinate (c1) at (rel axis cs:0,1);
      \nextgroupplot
        \pgfplotstableread[col sep=comma]
          {data/conv_visc_Cheb_N256_s0010_a1e-02.csv}\tabledataII
        \addplot table[x=k_CG, y=r_CG]{\tabledataII};
        \addplot table[x=k_BiCG, y=r_BiCG]{\tabledataII};
        \addplot table[x=k_BiCGStab, y=r_BiCGStab]{\tabledataII};
        \addplot table[x=k_RMR, y=r_RMR]{\tabledataII};
        \addplot table[x=k_GMRES, y=r_GMRES]{\tabledataII};
        \coordinate (c2) at (rel axis cs:1,1);
      \nextgroupplot
        \pgfplotstableread[col sep=comma]
          {data/conv_visc_Four_N256_s0100_a1e-02.csv}\tabledataIII;
        \addplot table[x=k_CG, y=r_CG]{\tabledataIII};
        \addplot table[x=k_BiCG, y=r_BiCG]{\tabledataIII};
        \addplot table[x=k_BiCGStab, y=r_BiCGStab]{\tabledataIII};
        \addplot table[x=k_RMR, y=r_RMR]{\tabledataIII};
        \addplot table[x=k_GMRES, y=r_GMRES]{\tabledataIII};
      \nextgroupplot
        \pgfplotstableread[col sep=comma]
          {data/conv_visc_Cheb_N256_s0100_a1e-02.csv}\tabledataIV;
        \addplot table[x=k_CG, y=r_CG]{\tabledataIV};
        \addplot table[x=k_BiCG, y=r_BiCG]{\tabledataIV};
        \addplot table[x=k_BiCGStab, y=r_BiCGStab]{\tabledataIV};
        \addplot table[x=k_RMR, y=r_RMR]{\tabledataIV};
        \addplot table[x=k_GMRES, y=r_GMRES]{\tabledataIV};
      \nextgroupplot
        \pgfplotstableread[col sep=comma]
          {data/conv_visc_Four_N256_s1000_a1e-02.csv}\tabledataV;
        \addplot table[x=k_CG, y=r_CG]{\tabledataV};
        \addplot table[x=k_BiCG, y=r_BiCG]{\tabledataV};
        \addplot table[x=k_BiCGStab, y=r_BiCGStab]{\tabledataV};
        \addplot table[x=k_RMR, y=r_RMR]{\tabledataV};
        \addplot table[x=k_GMRES, y=r_GMRES]{\tabledataV};
      \nextgroupplot
        \pgfplotstableread[col sep=comma]
          {data/conv_visc_Cheb_N256_s1000_a1e-02.csv}\tabledataVI;
        \addplot table[x=k_CG, y=r_CG]{\tabledataVI};
        \addplot table[x=k_BiCG, y=r_BiCG]{\tabledataVI};
        \addplot table[x=k_BiCGStab, y=r_BiCGStab]{\tabledataVI};
        \addplot table[x=k_RMR, y=r_RMR]{\tabledataVI};
        \addplot table[x=k_GMRES, y=r_GMRES]{\tabledataVI};
    \end{groupplot}
    \coordinate (c3) at ($(c1)!.5!(c2)$);
    \node[above] at (c3 |- current bounding box.north)
      {\pgfplotslegendfromname{fig:precond.visc.conv.vs_s.legend}};
    \tikzset{subcaption/.style={
      text width=6cm,
      yshift=-0.01\textheight,
      align=center,
      anchor=north}}
    \node[subcaption]
      at ($(plot group c1r1.south)!.5!(plot group c2r1.south)$)
      {(a) $a = 10^{-2},\, s = 10$};
    \node[subcaption]
      at ($(plot group c1r2.south)!.5!(plot group c2r2.south)$)
      {(b) $a = 10^{-2},\, s = 100$};
    \node[subcaption,yshift=-0.05\textheight]
      at ($(plot group c1r3.south)!.5!(plot group c2r3.south)$)
      {(c) $a = 10^{-2},\, s = 1000$};
  \end{tikzpicture}}{\includegraphics{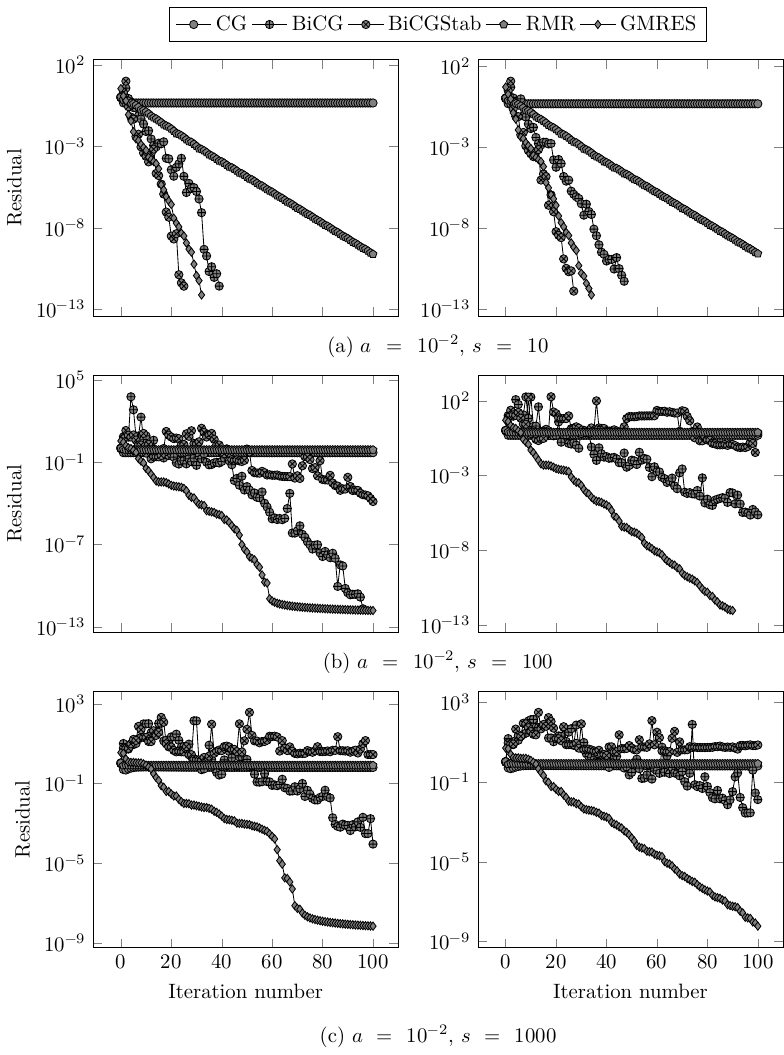}}
  \caption{\label{fig:precond.visc.conv.vs_s}%
    Convergence history for the five preconditioned iterative solvers
    to solve $\opVrho(\vec{u}) = \vec{b}$
    with $a = 10^{-2}$ and different values of $s$,
    with Fourier differentiation (left side)
    and Chebyshev differentiation (right side)
    and $N = 256$ collocation points.}
\end{figure}

\begin{figure}
  \centering
  \IfTikzLibraryLoaded{external}{
  \begin{tikzpicture}
    \begin{groupplot}[
      group style={
        group name=plot group,
        group size=1 by 3,
        ylabels at=edge left,
        xlabels at=edge bottom,
        x descriptions at=edge bottom,
        horizontal sep=0.1\textwidth,
      },
      width=0.7\textwidth,
      height=0.3\textheight,
      ymode={log},
      conv axis,
    ]
      \nextgroupplot[
        legend to name={fig:precond.pres.conv.vs_s.legend}
      ]
        \pgfplotstableread[col sep=comma]
          {data/conv_pres_Cheb_N256_s0010.csv}\tableDataI
        \addplot table[x=k_CG, y=r_CG]{\tableDataI};
        \addplot table[x=k_BiCG, y=r_BiCG]{\tableDataI};
        \addplot table[x=k_BiCGStab, y=r_BiCGStab]{\tableDataI};
        \addplot table[x=k_RMR, y=r_RMR]{\tableDataI};
        \addplot table[x=k_GMRES, y=r_GMRES]{\tableDataI};
        \addplot table[x=k_FP, y=r_FP]{\tableDataI};
        \legend{CG, BiCG, BiCGStab, RMR, GMRES, FP}
      \nextgroupplot
        \pgfplotstableread[col sep=comma]
          {data/conv_pres_Cheb_N256_s0100.csv}\tableDataII
        \addplot table[x=k_CG, y=r_CG]{\tableDataII};
        \addplot table[x=k_BiCG, y=r_BiCG]{\tableDataII};
        \addplot table[x=k_BiCGStab, y=r_BiCGStab]{\tableDataII};
        \addplot table[x=k_RMR, y=r_RMR]{\tableDataII};
        \addplot table[x=k_GMRES, y=r_GMRES]{\tableDataII};
        \addplot table[x=k_FP, y=r_FP]{\tableDataII};
      \nextgroupplot
        \pgfplotstableread[col sep=comma]
          {data/conv_pres_Cheb_N256_s1000.csv}\tableDataIII
        \addplot table[x=k_CG, y=r_CG]{\tableDataIII};
        \addplot table[x=k_BiCG, y=r_BiCG]{\tableDataIII};
        \addplot table[x=k_BiCGStab, y=r_BiCGStab]{\tableDataIII};
        \addplot table[x=k_RMR, y=r_RMR]{\tableDataIII};
        \addplot table[x=k_GMRES, y=r_GMRES]{\tableDataIII};
        \addplot table[x=k_FP, y=r_FP]{\tableDataIII};
    \end{groupplot}
    \node[above] at (current bounding box.north)
      {\pgfplotslegendfromname{fig:precond.pres.conv.vs_s.legend}};
    \tikzset{subcaption/.style={
      text width=6cm,
      yshift=-0.01\textheight,
      align=center,
      anchor=north}}
    \node[subcaption] at (plot group c1r1.south) {(a) $s = 10$};
    \node[subcaption] at (plot group c1r2.south) {(b) $s = 100$};
    \node[subcaption,yshift=-0.05\textheight]
      at (plot group c1r3.south) {(c) $s = 1000$};
  \end{tikzpicture}}{\includegraphics{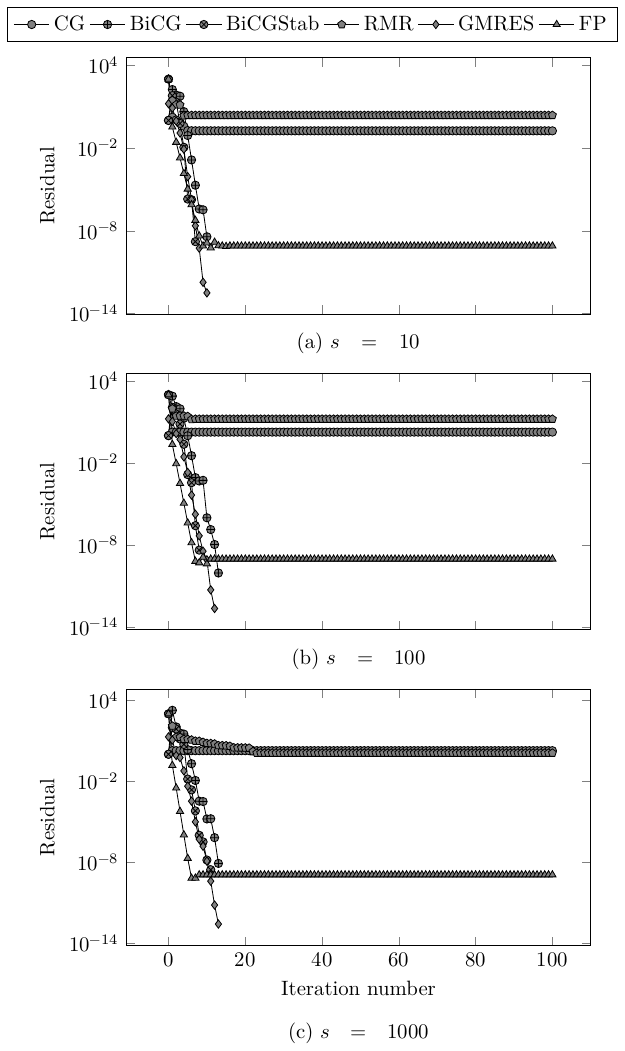}}
  \caption{\label{fig:precond.pres.conv.vs_s}%
    Convergence history of the different preconditioned iterative solvers
    to solve the modified pressure \cref{eq:VDINSE.Krho}
    with different values of $s$,
    and $N = 256$ Chebyshev collocation points.}
\end{figure}
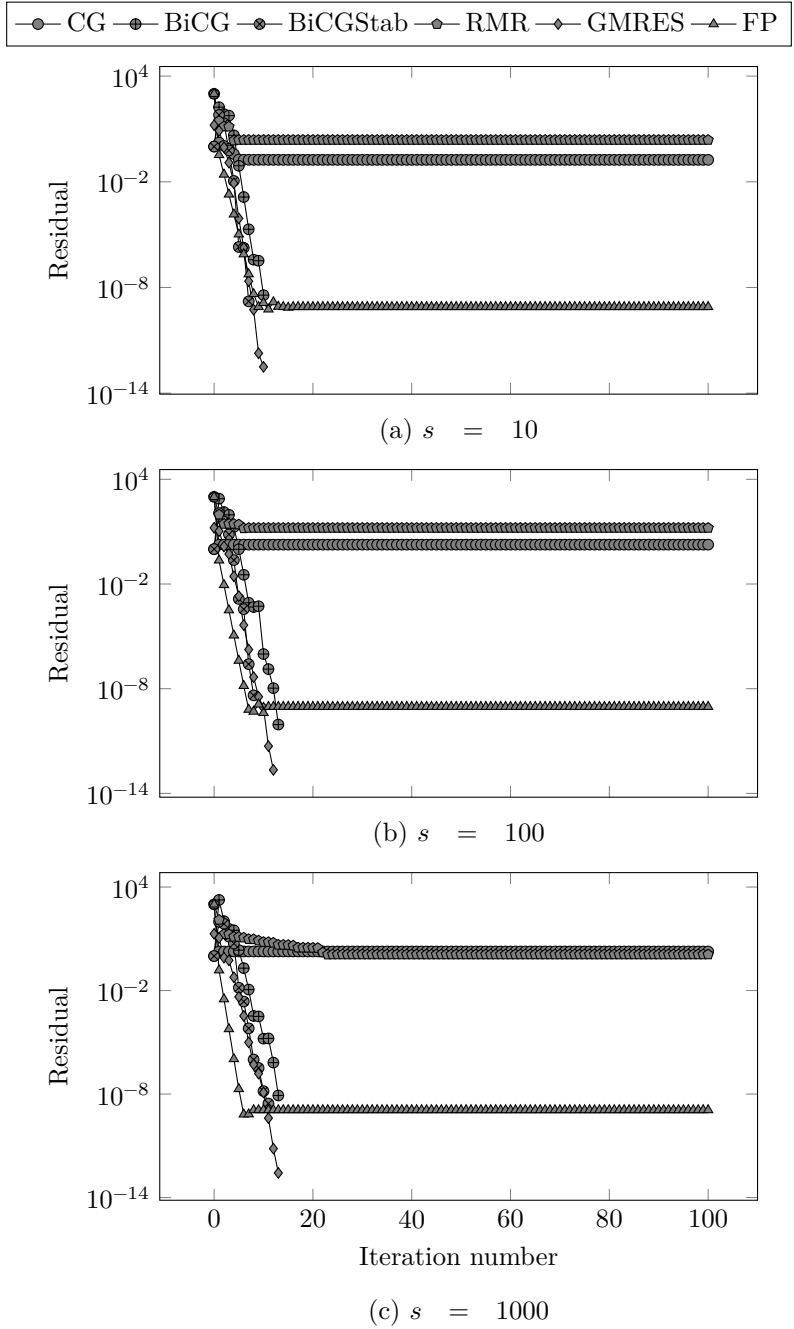

\bibliographystyle{elsarticle-num-names}
\bibliography{references}

\end{document}